# Simulation of Molecular Signaling in Blood Vessels: Software Design and Application to Atherogenesis

Luca Felicetti, Mauro Femminella, Gianluca Reali

{luca.felicetti, mauro.femminella, gianluca.reali}@diei.unipg.it

## Abstract

This paper presents a software platform, named BiNS2, able to simulate diffusion-based molecular communications with drift inside blood vessels. The contribution of the paper is twofold. First a detailed description of the simulator is given, under the software engineering point of view, by highlighting the innovations and optimizations introduced. Their introduction into the previous version of the BiNS simulator was needed to provide to functions for simulating molecular signaling and communication potentials inside bounded spaces. The second contribution consists of the analysis, carried out by using BiNS2, of a specific communication process happening inside blood vessels, the atherogenesis, which is the initial phase of the formation of atherosclerotic plaques, due to the abnormal signaling between platelets and endothelium. From a communication point of view, platelets act as mobile transmitters, endothelial cells are fixed receivers, sticky to the vessel walls, and the transmitted signal is made of bursts of molecules emitted by platelets. The simulator allows evaluating the channel latency and the footprint on the vessel wall of the transmitted signal as a function of the transmitter distance from the vessels wall, the signal strength, and the receiver sensitivity.

**Corresponding author:**
Mauro Femminella
DIEI - University of Perugia
via G. Duranti 93, 06125 Perugia, Italy
email: mauro.femminella@diei.unipg.it
ph: +39 075 585 3630, fax: +39 075 585 3654

# 1 Introduction

The capabilities of manipulating matter at the molecular scale has inspired a huge research effort for many years and has led to the design of sophisticated devices, commonly referred to as nanomachines. The potentials of these nano-devices span numerous areas [1][8][13], including medical science [2][14], environmental control, and material science. The technological progress has stimulated a lot of research focusing on different types of nanomachines, such as those related to the usage of carbon nanotubes and nanowires [51], and those making use of biological methods, which are the object of our analysis. Although full-fledged biological nanomachines do not yet exist, recently biochemists have done many progresses in this area, and now they are able to create at least functional cell components, such as artificial ribosomes, which can be used for the synthesis of proteins [49][50].

However, the research regarding nanomachine networked coordination is still at an early stage. In the last years, some possible solutions for allowing nanomachines to exchange information have been proposed [1]. Clearly, achieving the objective of exchanging information at the nanoscale requires a deep exploration of the feasible mechanisms that allow designing the basic components of a communications system, such as an information encoder, a transmitter, a communication medium, a receiver, and an information decoder [13].

Due to the heterogeneity of different environments and communication techniques that can be used at the nanoscale, it is unfeasible to identify general models, valid for most of nano-communication systems. For example, signaling within a lymph node, or within blood vessels, or between brain cells, make use of mechanisms specific for each environment, different from each other. Hence, their analysis require different models, strictly related to their environmental features. For this reason, through the combination of interdisciplinary expertise, for each specific environment that could host nanoscales communications, it is necessary to plan and execute suitable experiments, which can be both in laboratory and in silico, in order to achieve a deep knowledge of it.

The main objective of our research is to analyze diffusion-based molecular signaling in blood vessels, modeled as molecular communications by means of the upgraded version of the BiNS simulator [3], which a simulation package for biological nano-scale communications developed at the University of Perugia.

In more detail, the contribution of this paper is twofold. The first one consists of the detailed presentation of BiNS2, including its basic processing mechanisms, from a software engineering viewpoint, by highlighting newly introduced functions needed to simulate molecular communications inside blood vessels. In fact, in order to correctly simulate molecular signaling, based on diffusion with drift in bloodstream, it has been necessary to model (i) the presence of bounding surfaces representing the vessel walls, (ii) the movement of particles (white and red blood cells, platelets, and emitted molecules) taking into account the brownian motion, the drift effect due to the blood pressure, and collisions, and (iii) the exchange of blood cells with the exterior of the simulated vessel section, as a consequence of the blood flow. In addition, BiNS2 can be executed in a distributed environment, by using a computation grid approach.

The objective of our research is also medical. In fact, the second contribution of this paper is the presentation of the results achieved by using BiNS2 to simulate a significant case study in the medical area. It consists of the molecular-based signaling, in blood vessels, between platelets and endothelium, which is widely recognized as the initial phase of the cardiovascular disease known as atherosclerosis, i.e. the atherogenesis [19][21][22]. In the simulated system, the platelets act as mobile signal transmitters, the endothelial cells act as fixed receivers, and the signal is made of bursts of special molecules, the sCD40L cytokines [18], which are carriers released by the transmitters. The presence of white and red blood cells contributes to "shape" the signal since, being their size much larger than the carrier size, collision with them push carriers towards the vessel walls. In addition, since white blood cells are also able to absorb carriers, they contribute to the signal attenuation. Clearly, through this analysis we can also gain more insights from a communications engineering point of view about the critical process of atherogenesis. To this aim, we have also executed a number of biological lab experiments, the main findings of which are illustrated in [5][37]. The objective of these experiments was not simply to observe the outcome of the investigated biological mechanisms, which was already known, but rather their evolution. This way, through the knowledge of the transient behavior of the involved biological interactions, we have designed, implemented, and tuned suitable emulating procedures in the BiNS2 package. Clearly, experiment planning and simulator development have a mutual influence on each other, since any simulation achievement contributes to improve understanding and is useful to direct future experimental activities on the investigated aspects.

This research also contributes to the further technical objective of identifying mechanisms enabling generic nanomachines communications in the bloodstream environment, using molecular communications, without limiting our interest to the analysis of the interactions between platelets and endothelium through sCD40L molecules.

These research objectives contribute to the longer term goal of designing and realizing nanomachines (nano-sensors/actuators) for the prevention and treatment of cardiovascular diseases, also through the interaction with the outer world, that is artificial devices located out of the human body able to exchange information with nanomachines, as sketched in [4].

The paper is organized as follows. In section 2 we illustrate the background and related work in the field. In section 3 we illustrate our simulator and its core functions for emulating communications in blood vessels. In section 4, we both illustrate the specific scenario we have analyzed and show the results of our simulations. Finally, in section 5, we draw some concluding remarks.

## 2  Background and Related Works

In this section, we first illustrate the related work on the nano-networking reseach, by describing in some details the contributions dealing with diffusion-based molecular communications, and the relevant simulation platforms. The second subsection settles the background for the use case (i.e. the study of the initial phase of atherogenesis), which will be analyzed in section 4 by using the BiNS2 simulator described in section 3.

### 2.1  Related Works on Nanonetworks

In this section, we review the achievements in the field of nano-scale communications in biological environments. Typically, molecules are used as information signal and their diffusion is the mechanism governing the signal propagation, often modeled as a Brownian motion [24]. In [7] mathematical models of transmitter, channel, and receiver are shown. Nodes are assumed to be fixed. The authors evaluate the end-to-end gain and propagation delay as a function of some environmental parameters. Information is transferred by modulating the concentration of the carriers (molecules) emitted by the transmitter. In the same scenario, papers [29] and [30] analyze the noise sources affecting the diffusion-based molecular communication. A slightly different scenario is depicted in [32], where the transmitter is placed within a fluid environment and emits a series of identical molecules, which disperse in the environment by a Brownian motion and are absorbed by a receiver capable of taking into account their arrival times. In this scenario, information is encoded in the release time of molecules as in a pulse position modulation, instead of encoding it in their concentration.

Some research effort have focused on evaluating the channels capacity in diffusion-based nano-networks. In [31] a stochastic model for molecular reactions in biochemical systems, and the relevant channel model, is shown. On the basis of this model, the authors provide a deterministic capacity expression for point-to-point, broadcast, and multiple-access molecular channels. They also evaluate the information flow capacity in a molecular nanonetwork made of fixed nodes. In [6], the authors propose a slightly different scenario, in which both nano-nodes and emitted molecules propagate through a fluid medium, propelled by a drift velocity and Brownian motion. They analyze a preliminary model of a communication system based on the release of either one or two molecules into the fluid medium. In [33] the same authors show that the additive inverse Gaussian noise channel is appropriate for modeling molecular communication channels in fluid media with drift. They also derived upper and lower bounds on channel capacity and a maximum likelihood receiver.

In [37] formal models, based on finite state machine diagrams, are applied to inter and intra-cells interactions.

Other works focus on computer simulations of biological nanonetworks. In [9] the authors present a simulator based on NS-2, able to simulate the Brownian diffusion within a tridimensional environment. Another approach, making use of a Java platform, implements the Brownian diffusion process within a two dimensional space. The activity for introducing a tridimensional extension, under specific scenarios [10], is ongoing. Finally, it is worth citing the BiNS simulator [3], which is the platform used in this work to simulate communications in blood vessels. In principle, also multi-physics simulation platforms, non specific to nano-communications, could be used to simulate nanoscale environments, such as the well-known COMSOL Multiphysics® simulator [25]. Nevertheless, these platforms are commonly used to model phenomena at a macroscopic level, such as flows, and not the interaction of single particles. When used to model more detailed interactions, such as particle tracing in fluids, they need of specific models. In particular, no specific libraries are available for simulating the interactions of biological entities, such as the ligand-reception formation, and specific libraries have to be implemented by users. In addition, the implementation of the extensions of these generic platforms, needed for modeling specific communications phenomena in molecular communications, such as the inter-symbol interference [53], or for implements signal processing modules, such as the Viterbi algorithm [53], seems to be more difficult than introducing them within an open simulation platform like BiNS2.

## 2.2 Biological Background on Atherogenesis

The biological background of this paper regards the interaction between the platelets and the endothelial cells in blood vessels during the atherogenesis. It is still controversial which is the initial trigger of atherogenesis. A hypothesis is that it is caused by an anomalous behavior of endothelium, which recruits platelets on the injury site through tissue factor production and subsequent release of thrombin [16]. Another hypothesis is that it is due to an anomalous activation of platelets, since their interaction is critical for the initial phases and subsequent development of atherosclerosis [17][19]. In their interaction with the endothelium, platelet cells behave as mobile transmitter nodes, and the endothelial cells as fixed receiver nodes. As mentioned above, the signal transmission in this communication environment consists of the release of a special kind of cytokines, acting as messengers, called CD40L (also known as CD154), which is a trimeric, transmembrane protein of the tumor necrosis factor family [18][21][22]. Resting platelets store CD40L inside the cytoplasm, but do not express it on the cells surface. Upon receipt of external stimulus (thrombin), platelets activates and express CD40L on the cell surface. The CD40L expressed on platelet surface is subsequently cleaved and shed from the surface as sCD40L (soluble CD40L). Platelets can be deactivated by using hirudin [37]. As for the receivers, endothelial cells express the corresponding CD40 receptors on their surface. Upon CD40 receptors on the endothelial cell surface begins binding to sCD40L in blood or to CD40L on platelet surface due to mechanical contact, endothelial cells start expressing additional CD40 receptors, and receptor-ligand bindings are internalized [20]. If the stimulus intensity is strong and permanent enough, the trafficking process in endothelial cells (which is the intracellular process consisting of bindings internalization, signaling, and subsequent receptor recycling, or resynthesis) [23] starts producing vascular cell adhesion molecules (VCAM-1) on the surface of cells (endothelium activation). These adhesion molecules cause monocytes to adhere to endothelium. Since also monocytes express CD40L, a contact with CD40 receptors on endothelium reinforces VCAM-1 production. Finally, after their recruitment, monocytes start penetrating below endothelium towards the inflammatory site (diapedesis process).

This kind of communication and interaction happens when blood vessels are injured. In case of abnormal behavior, this process leads to monocytes transformation in macrophages, which in turn become foam cells, which constitute the initial elements of the fatty streak in the atherosclerotic plaques. Instead, in normal conditions, platelets are not activated and do not interact with the endothelium, flowing along the blood vessel in the proximity of the vessel walls, due to collisions with other bigger cells (mainly red blood cells) that propagate with them and happen to push them away towards vessel walls [15].

Research on molecular communications finds a straightforward cross-fertilization with important ongoing research activities in the biological and medical fields involving nano-machines. For exemple, [55][56][57] illustrates an ongoing research activity related to the use of nanomachines for sensing the formation of atherosclerotic plaques and their stabilization. In this case the biological mechanisms analyzed and exploited are the same involved in our research. Thus, the long term objectives of our research illustrated in Section 4 are complementary to the research activity shown in [55][56][57], since the same biological interactions may be used for both communicating specific plaque fixing actions and monitor their evolution.

## 3  The Software Library

In this section, we present an overview of the functions available in the BiNS2 simulator. Section 3.1 briefly illustrate the general features of BiNS2, in order to provide the reader with the information essential to appreciate the contribution of this paper. The interested reader can find additional details in [3].

Section 3.2 illustrates the usage of domains, an architectural improvement of the simulator that allows simulating more complex scenarios, such as bounded spaces. Finally, Section 3.3 reports how BiNS2 can benefit of grid computation techniques in terms of simulation scale and execution time.

### 3.1  Simulator Structure

The BiNS simulator [3] has been designed with the aim of making it highly customizable. A set of tools is available. It allows creating software objects modeling the behavior of biological entities, which can be regarded as either nodes (transmitters, receivers, or both) or carriers, which are emitted by transmitters and constitute the signal in molecular-based communications. In addition, it is possible to configure the properties of the simulated communication channel (e.g. the blood stream or the environment of an in vitro experiment) with the desired accuracy.

Each object belongs to a generic type of software object, named Nano Object. Nodes and carriers are specific implementations of the Nano Object, and, although they share its general features, they can expose very different functions. In particular, for any scenario, it is possible to differentiate the node object type at any time, thus obtaining a multitude of different node objects. The same properties are available for modeling carriers as Nano Objects.

The simulation is organized in discrete time steps. Each step consists of a number of phases, in which software objects are triggered in order to execute the operations associated with their specific behavior. The main phases are:

- transmission phase, i.e. emission of carriers;
- reception phase, i.e. carriers assimilation;
- information processing phase, which is executed depending on the received signal intensity, which corresponds to the number of assimilated carriers;
- motion phase, during which all mobile objects (node and carriers) are moved according to the rules specified for them in the simulation configuration (mobility model);
- object destruction phase, during which objects are removed due to lifetime expiration or because they exited from the area of interest;
- collision management phase, which, in turn, consists of a collision check phase, during which the objects under collision are identified, and a relocation phase, during which those objects are moved as a result of the (in)elastic collision.

Clearly, not all types of objects will execute all phases. For instance, a carrier does not execute any transmission, or a fixed object does not execute the motion phase.

The simulator uses a fine grained approach for handling collisions between objects. A collision can produce either a bounce or an assimilation. The latter happens only when a carrier collides with a compliant receptor on the surface of a node.

The simulated environment can be either unbounded or bounded by a surface of custom shape. For instance, in section 4 we show that in order to simulate communications within a blood vessel, we have decided to limit the simulation scope to a cylindrical volume. This new feature has been made available in BiNS2, and described in detail in Section 3.2.

The mote computational intensive simulation phases, such as the motion and collision management phases, are handled through parallel computing techniques. The whole list of nano objects of the simulation are split into smaller lists, which can be handled in parallel by a thread pool, by taking into account concurrency and synchronization issues. Clearly, this approach is more effective in multicore platforms, since these systems can efficiently execute a large number of parallel threads, and thus it is possible to crease sublists with a smaller number of objects to be managed. Important issues of this approach are the determination of the suitable number of threads and the workload distribution to them, in order to avoid that a single thread could slow down the execution of all the other threads in the pool during a time step. In addition, it is also necessary to consider the overhead associated with the management of each thread. The optimal configuration was found experimentally, since it depends on the hardware of the used servers. The rule of thumb that has emerged suggests to create a number of threads larger by 4 to 8 times than the number of available CPU cores.

## 3.2 Introduction of Custom Domains

In order to model different scenarios of molecular communications more accurately, we have introduced a generalized approach to deal with the spatial domains. The proposed approach is hierarchical, and splits the overall simulated environment into smaller volumes of well-defined geometrical shapes. It is possible to define 4 domain types, which differ by their shape: spherical, cubic, cylindrical, and unbounded.

Each domain has the knowledge of all nano objects contained in it, implements the mobility model that defines the motion rules inside its managed space, checks when an object has to be destroyed, and manages the collision handling. The latter can be changed even at runtime in order to switch between different collisions sensing/handling strategies relevant to different environments in the same simulation. Hence, each domain can handle its own space and all embedded objects independently. This feature is useful for simulating different parts of the same experiment, also involving different regions with specific propagation laws. As a result, the introduction of domains separates the spatial features from the management features, such as the initialization of the simulation scenario, the management of the discrete time steps, and the collection of results.

Domains we could model even mobile objects, for instance to model what happens *inside* a nanomachine or a cell, if necessary. In this case, the domain would manage the position of (eventually) sub-domains and nano objects embedded inside the modeled entity. This feature is not used in the simulation scenario described in Section 4.

The most important task in terms of used CPU time assigned to domains is the management of collisions. The algorithm used to manage collisions of nano objects, when domains are present, need a more detailed explanation, since it represents one of the main novelties of BiNS2. In fact, the introduction of domains has made necessary to differentiate the type of collisions of nano objects. So, it is needed to handle differently collisions between two nano objects (Figure 1.a) and collisions between a nano object and a domain boundary, regardless the type of collision occurred, which can be with the inner domain surface (Figure 1.b) or with the outer domain surface (Figure 1.c). In the cases illustrated in Figure 1.b and Figure 1.c, it is necessary to take into account the domain shape. The detection of collisions involving domains is explained in more details in Section 3.2.1.

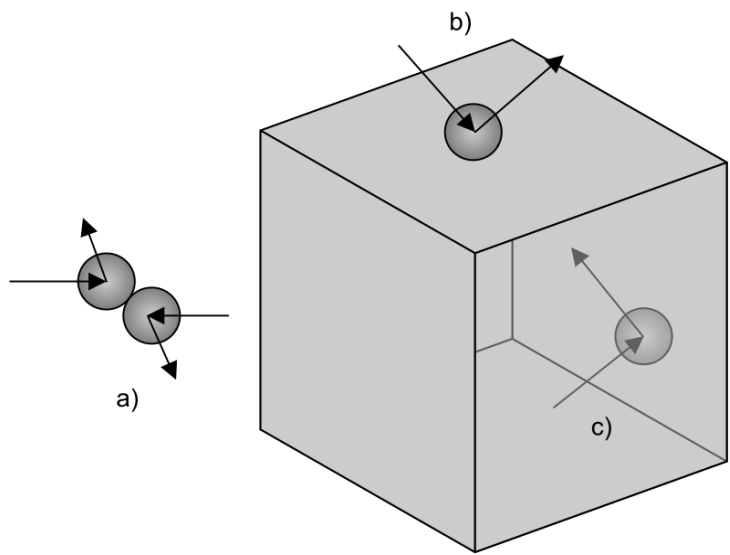

Figure 1 - Different types of collisions. a) between nano objects, b) external collision, c) internal collision.

Each domain is hierarchically attached to its parent domain, thus its position in the 3D space is defined as a function of the coordinates of its parent domain, as shown in Figure 2. At the highest level of the hierarchy, there is the Root domain, which is a special kind of unbounded domain.

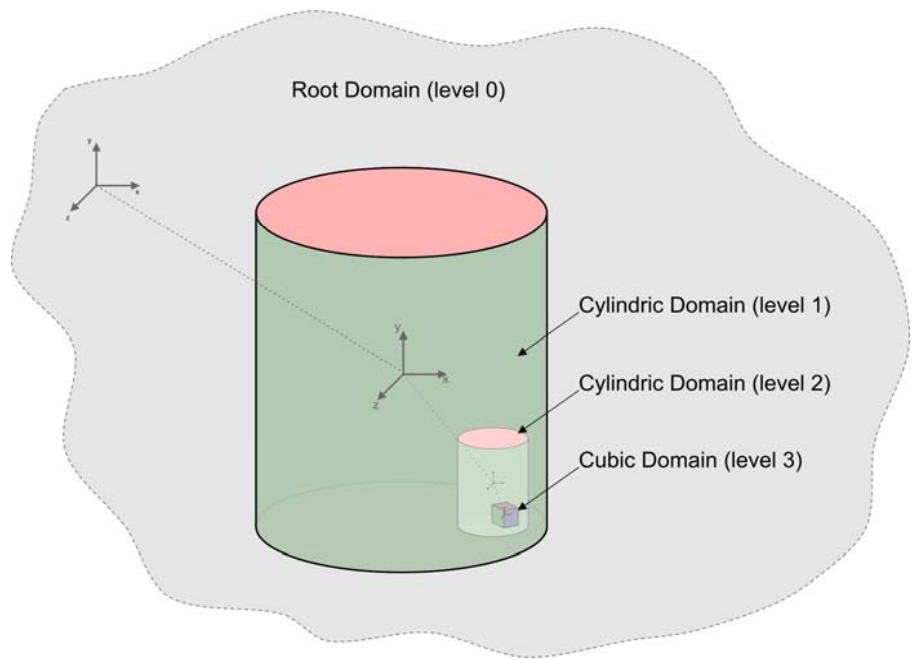

Figure 2 - Hierarchical management of system of coordinates in nested domains.

Each inner domain is compared with any nano object that belongs to its parent domain by using an algorithm derived from the one adopted to find out collisions between nano objects. For the sake of clarity, we briefly recall the two main phases of this algorithm; additional details can be found in [3]. In the first phase, objects are ordered in a list according to the distance of their center from an arbitrary reference point in descending order (i.e. from the farthest to the nearest). In the second phase, collisions are checked by comparing *only* the adjacent objects in the list in order to minimize the computational effort. This way of handling collisions reduces the asymptotic complexity from O($n^2$) to the $O(n\log(n))$. This means we are switching from a three-dimensional (3D) reference system to a mono-dimensional (1D) reference system. This allows implementing a preliminary, simplified check, since if collisions do not occur in the 1D system, they will not occur in the 3D one. Instead, if a collision occurs in the 1D system, it may not occur in the 3D system. Thus, a verification in the 3D system has to be done anyway, but *only* for the pairs of objects which are detected as "colliding" in 1D system by the second phase of the above algorithm, thus strongly reducing the number of 3D checks.

When different domains are used, the simulation is speedup , since handling many small sublists is faster than handling a large one. In fact, if the total number of nano objects is *N* and it is split into *D* domains, in the case of equally distributed nano objects (best case), any domain will have to manage *M=N/D* nano objects. Then, the complexity of the collision checks will be

$$O(M \cdot D \log(M)) < O(N \log(N)) \qquad (1)$$

if the checks were managed sequentially. Since all of these steps can be executed in parallel threads, it could be possible to reach the theoretical limit of $O(M \log(M))$ by deploying *D* parallel threads, if the number of available CPU cores is no lower than *D*. Obviously, the overhead due to the management of multiple threads, concurrency issues, and their synchronization have an significant impact on the achievable performance and has to be taken into account.

The following sections show how collisions are managed when domains are used. In more detail, Section 3.2.1 describes the algorithm implemented to manage the collisions between nano objects and inner domains contained within a parent domain. Section 3.2.2 describes the algorithm used to detect the collisions between nano objects and walls of the domain. Finally, Section 3.2.3 describes the algorithm used to manage the relocation of nano objects after a collision with a wall of the domain. The algorithms in Section 3.2.2 and Section 3.2.3 are specialized for the cylindrical domain, which is the domain type used in the simulation of a blood vessel described in Section 4.

### 3.2.1 Collisions Detection with Inner Domains

The collisions detection between two or more nano objects having a spherical shape has been briefly summarized above. Instead, the collision handling between a nano object and the wall of a domain has to consider the domain-specific shape. In order to handle this task and reuse the same approach designed for detecting the collision between nano objects, each domain is associated with a bounding sphere. It is defined as the sphere with the minimum radius that contains the whole domain. We have designed and implemented a preliminary check which makes use of the bounding sphere of the domain, in order to compare two spherical objects, which is easier than using directly a domain with an arbitrary shape. Since the bounding sphere contains the domain, this approach may detect some false collisions. However, these false positive will be identified and correctly managed on the subsequent phase by a more accurate, but slower, control algorithm, specifically implemented for each specific domain shape. This preliminary check is executed after the check of collisions between any couple of nano objects.

Thus, any domain has to control if each contained nano object collides with each internal domain. As shown in Figure 2, the reader has to remember that the reference point for collision check is the center of the domain handling the check. First, the internal domains are sorted from the farthest to the closest to the reference point, so resorting to the 1D system of coordinates described in Section 3.1 and illustrated in Figure 3. Then, each couple composed by one inner domain and one nano object is evaluated, in a similar way as illustrated in the previous section for pairs of nano objects.

As shown in Figure 3.a, let $C_d$ be the coordinate value in the new 1D system of the center of the considered inner domain, and let $D_-$ and $D_+$ be defined as follows, with $R$ the radius of the inner domain:

$$D_- = C_d - R \qquad (2)$$

$$D_+ = C_d + R \qquad (3)$$

In addition, let $C_n$ be the coordinate value of the center of the nano object under consideration for the collision check with the inner domain above, and $n_-$ and $n_+$ be defined as follows, with $r$ the radius of the nano-object:

$$n_- = C_n - r \qquad (4)$$

$$n_+ = C_n + r \qquad (5)$$

Now, let us consider the following three cases:

1. If $D_- > n_+$ then there is no overlapping between the current domain and any previous nano object. The checks will continue with the next (dashed) domain, as shown in Figure 3.b.
2. If any of the following conditions

$$D_+ \leq n_- \leq D_- \qquad (6)$$

$$n_- \leq D_- \leq n_+ \qquad (7)$$

$$n_- \leq D_+ \leq n_+ \qquad (8)$$

   is valid, then a collision *may* happen (see Figure 3.c). A more accurate control will be delegated to the specific collision strategy class associated to the current domain.
3. No collision happens between the current domain and the current nano object (red colored in Figure 3.d). The control will continue with the previous nano objects.

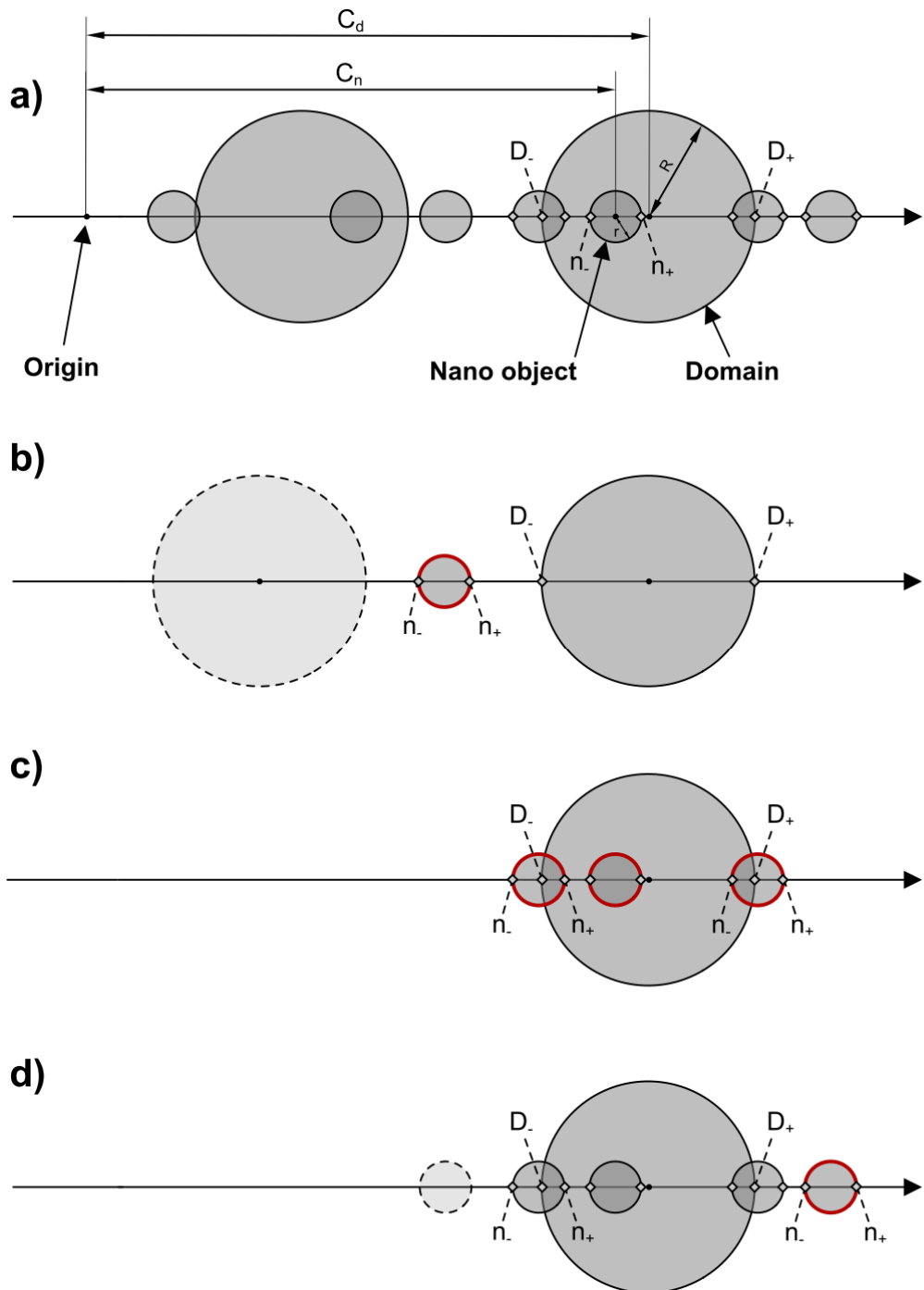

Figure 3 - Preliminary check collision between inner domains and nano objects.

### 3.2.2 Detection of Collisions with Cylindrical Domains

We now describe the algorithm implementing the collision detection in a specialized domain, the cylindrical domain. It has a special importance in this paper, being it a realistic model for blood vessels. We describe the case of collisions with the inner surface of the cylindrical domain and explain the model of the impact between nano objects (carriers and blood cells) and the vessel wall.

The internal collision may happen in two ways: (i) with either the top or bottom surfaces, or (ii) with the side surface. The need of modeling an object collision with the top and bottom surfaces is due not only for representing particular situations in blood vessels, such as an obstruction, but also for emulating lab biological experiments, such as the one described in [37]. Clearly, when a section of regular blood vessels are emulated, top and bottom surfaces absorbs objects, without bounces. The side surface of the cylinder is not permeable, that is the nano objects cannot pass through it, since in the case analyzed in this paper the cylindrical domain models a blood vessel.

Let $\hat{d}$ be the unit vector of the longitudinal axis of the cylinder and $\hat{n}$ the unit vector of an axis orthogonal to $\hat{d}$ such that $\hat{n}$ is rigidly coupled with the colliding nano objects (i.e. $\hat{n}$ is the unit vector which identifies the axis orthogonal to $\hat{d}$ and passing through the center of the nano object), as shown in Figure 4.

Let $\vec{C}_{n1}$ and $\vec{C}_{n2}$ denote the distance vector from the center of the cylinder of a nano object that collides with the top surface and with the side surface, respectively. The projections of $\vec{C}_{n1}$ and $\vec{C}_{n2}$ on the axes $\hat{d}$ and $\hat{n}$ are:

$$d_{p1} = \vec{C}_{n1} \cdot \hat{d} = |\vec{C}_{n1}|\cos(\beta) \tag{9}$$

$$d_{n1} = |\vec{C}_{n1}|\sin(\beta) \tag{10}$$

$$d_{n2} = \vec{C}_{n2} \cdot \hat{n} = |\vec{C}_{n2}|\cos(\alpha) \tag{11}$$

$$d_{p2} = |\vec{C}_{n2}|\sin(\alpha). \tag{12}$$

Let $h_v$ and $r_v$ be the cylinder height and radius, respectively, and $r_1$ and $r_2$ the radius of nano objects $N_1$ and $N_2$ in Figure 4, respectively. A collision happens if the following conditions on the projections on the axes $\hat{d}$ and $\hat{n}$ are verified. For what concerns a collision with the top surface the condition is

$$d_{p1} + r_1 \geq \frac{h_v}{2}, \tag{13}$$

whereas the condition for a collision with the side surface is

$$d_{n2} + r_2 \geq r_v \tag{14}$$

Figure 4 provides a graphical interpretation of these formulas.

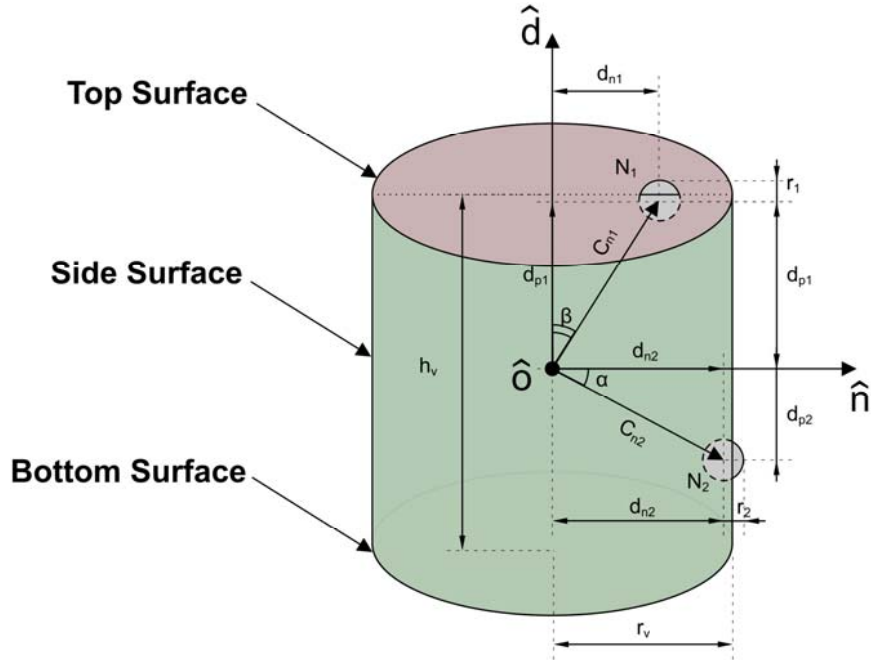

Figure 4 - Collisions detection on the cylindrical domain.

### 3.2.3 Relocation of Nano Objects within the Cylindrical Domain

Each detected collision is handled in order to bounce the nano object off the inner surface of the cylinder, depending on its specific shape (flat or curve surface). For both cases, the collision handling procedure checks for any overlap between the nano object and the considered surface at the end of the time step so as to calculate the average speed in that time interval, which, in turn, is used to evaluate the time of impact, similarly as in [3].

First, we focus on collisions with the top or bottom surfaces. Let us define the unit vectors of the three axes of the cylinder as $\hat{d}$, $\hat{n}$, and $\hat{o}$, respectively; the projections of the velocity vector ($\vec{V}_{ni}$) of the considered nano object on these axes (see Figure 5) $\vec{d}_i$, $\vec{n}_i$, and $\vec{o}_i$, are given by:

$$\vec{d}_i = \hat{d}(\hat{d} \cdot \vec{V}_{ni}) \tag{15}$$

$$\vec{n}_i = \hat{n}(\hat{n} \cdot \vec{V}_{ni}) \tag{16}$$

$$\vec{o}_i = \hat{o}(\hat{o} \cdot \vec{V}_{ni}) \tag{17}$$

The final velocity vector $\vec{V}_{nf}$ after the bounce, in case of elastic collision, is given by:

$$\vec{V}_{nf} = \vec{n}_i + \vec{o}_i - e\vec{d}_i, \tag{18}$$

where $e$ is the coefficient of restitution, used to model the inelastic collisions [54]. It ranges from 0 (completely inelastic collision) to 1 (elastic collision). In order to model the inelastic collision, only the component which is orthogonal to the colliding surface has to be damped by $e$. The relevant mathematical details can be found in the Appendix. Clearly, if the final velocity $\vec{V}_{nf}$ has to be determined by an inelastic collision, then $\vec{V}_{nf}$ is a fraction of the initial velocity $\vec{V}_{ni}$.

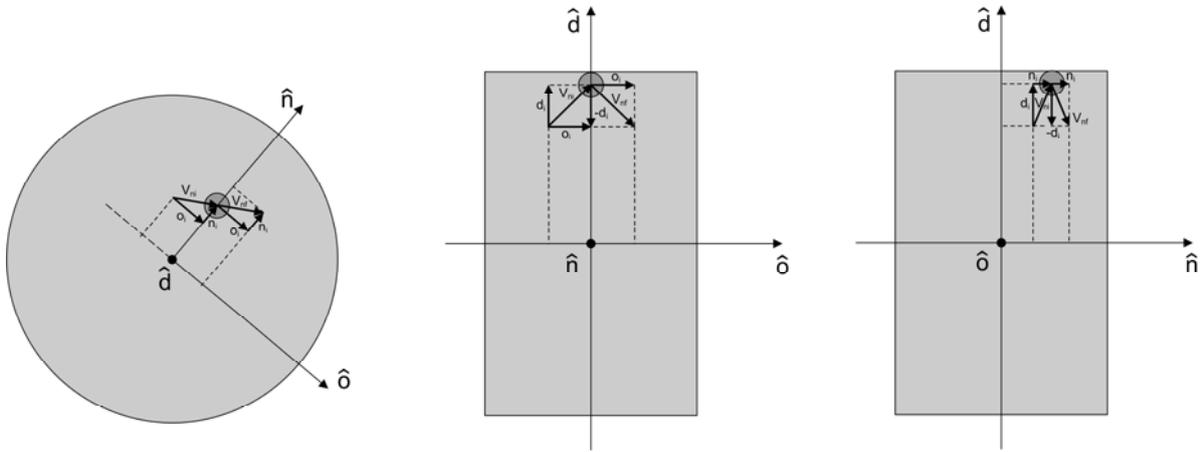

Figure 5 - Collisions handling on top surface.

The second case regards a nano object that collides with the side surface of the cylinder, illustrated in Figure 6. It is handled similarly as in the previous case, by projecting the velocity vector of the colliding nano object on the axes $\hat{d}$, $\hat{n}$, and $\hat{o}$. The resulting velocity vector is given by:

$$\vec{V}_{nf} = \vec{d}_i + \vec{o}_i - e\vec{n}_i. \tag{19}$$

In the case the surface involved by the collision is the side one, the only component dumped by $e$ is $\vec{n}_i$.

In the scenario considered in this paper, that is the object movement inside a blood vessel, a suitable value for the vessel wall restitution coefficient $e$ is 0.6 [11].

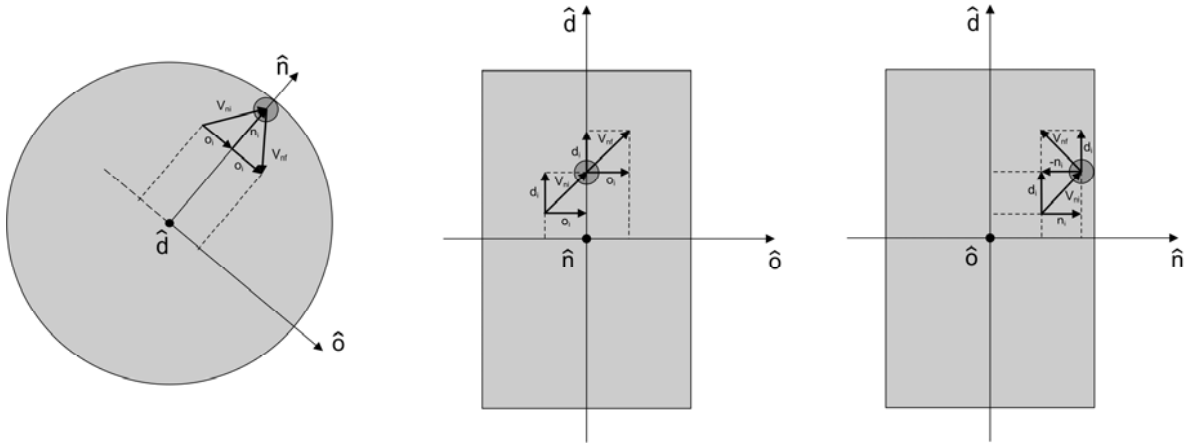

Figure 6 - Collisions handling on side surface.

### 3.3 Distributed Implementation of BiNS2

In the new version of the simulator we have introduced the possibility to run the BiNS2 simulator over a computational grid infrastructure using the well-known GridGain framework [40]. In order to enable this feature, the overall simulation space was partitioned into a number of smaller volumes, each managed by a different grid node, either physical of virtual. Thus, we split the simulation into a number of parallel simulations, each in charge to manage a portion of the split volume. These sub-simulations complete all phases associated to each time step independently (see Section 3.1). Nevertheless, they have to be synchronized before proceeding with the following time step. In more detail, each grid node has to wait a synchronization signal before proceeding with the subsequent phase. This way it is possible to correctly handle any possible exchange of moving nano objects between grid nodes handling neighboring volumes (i.e. volumes originated by the splitting process and having a common wall) and to prevent any incongruence.

This new feature introduces several benefits, such as the possibility of both executing larger simulations and sharing computational resources. Nevertheless, it also introduces some performance issues. The first one is the need of using a splitting algorithm able to efficiently split the simulation volume in different simulation environments, and not tailored to just a specific case study. In the current version of BiNS2, this algorithm is optimized for simulating blood vessels. A further issue is the node synchronization, which introduces wait times that can affect the simulation time. Finally, also the transfer of nano objects between different grid nodes due either collisions or nano object movement, may introduce a significant overhead, since any object has to carry along all its references (e.g. to the current domain). These references have to be restored at the destination grid node, thus resolving any redundancy.

We have evaluated experimentally that, if the grid nodes have a CPU utilization below 35% of its capacity, the synchronization management introduces a fixed delay which makes it useless for small simulations. Thus, it results to be useful only for simulations involving tens of thousands or millions of nano objects. In this way, for each time step, most of the time is spent in the simulation phases and not just for waiting synchronization signals, thus taking benefits from both the aggregated memory size and computational resources. In what follows we present the main features of the grid implementation of BiNS2.

For ease of handling, the elementary volume used to split the overall environment is a cube, which is a common choice when a volume has to be split into different sub-volumes [38] to perform parallel computations, thanks to the symmetry of this geometrical figure.

All resulting cubes have the same size and contain only the relevant portion of the simulation environment. Each node managing a cube is also aware of the six neighboring cubes. This awareness simplifies not only the splitting algorithm but also the communication mechanisms between cubes. We have implemented a conventional adjacency order by tagging the six faces of the cube as a playing dice, as shown in Figure 7. In the current implementation, the splitting algorithm is executed once at the beginning of the simulation. Thus, it is a static splitting, differently from what can be done, for instance, using octree-based algorithms [38]. This choice does not imply degraded performance, but higher programming effort just for the simulation setup, since it is not fully automated. In future versions we will improve this feature. The total number of cubes is decided by the user, according to the estimated simulation needs. For any axis, only an integer number of cubes can be disposed. Thus, the splitting algorithm has to calculate the positions and the adjacency lists of the cubes, as sketched in Figure 8. These positions are calculated by using the center of the system of coordinates of the original simulated volume illustrated in Figure 2. The cube positions are relative to this point. For example, if we consider the scenario shown in Figure 8, denoting the center of the system of coordinates as [$x_0$, $y_0$, $z_0$] and assuming a length of the side of each cube equal to $2a$:

- Cube #1: center on [$x_0 - a$, $y_0 + a$, $z_0$]
- Cube #2: center on [$x_0 - a$, $y_0 - a$, $z_0$]
- Cube #3: center on [$x_0 + a$, $y_0 - a$, $z_0$]

- Cube #4: center on $[x_0+a, y_0+a, z_0]$

Once the splitting is done, each portion of the space can be assigned to one of the grid nodes. Each node can, in turn, start its own simulation inside the managed cube for parallelizing the execution.

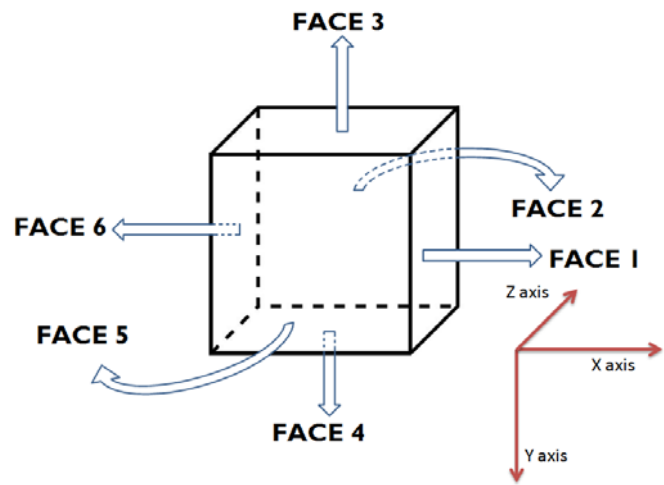

Figure 7 - Cube domain adjacencies.

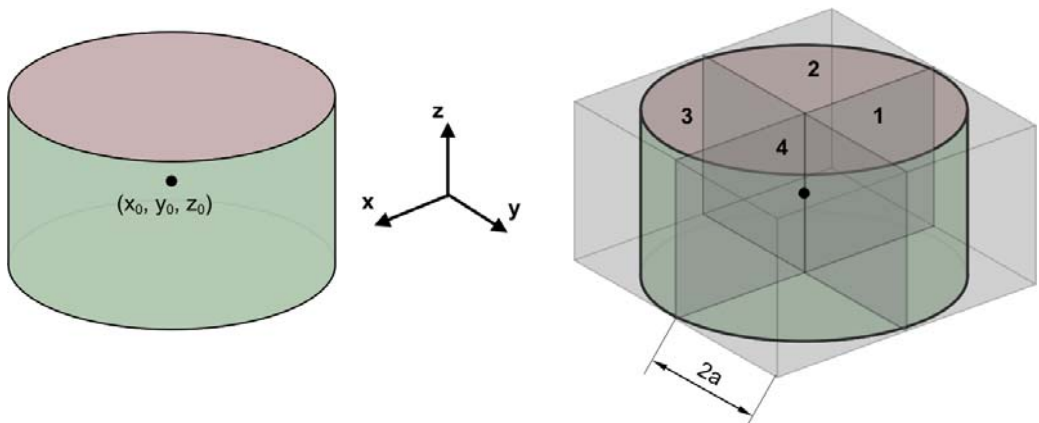

Figure 8 - The splitting strategy.

All parallel sub-simulations must be synchronized, otherwise it would be impossible to correctly execute basic operations, such as collision detection and output saving. The synchronization procedure is the performance bottleneck of our distributed computing environment, since every time it is executed, the individual processing must be stopped and grid nodes enter an idle state, until all nodes reach the same state of the computation. The execution time of each synchronization phase is indicated by a clock in Figure 9. The number of synchronization phases required for each time step is 2, occurring after an object is transferred from a grid node to another one (it happens in the object transfer block in Figure 9), by using the distributed cache function provided by GridGain [40]. In fact, it is necessary to synchronize grid nodes for checking for any incoming/outgoing objects for every node when they are moved. This happens after any object movement phase and after object relocation upon collisions. The feedback to the object transfer blocks in Figure 9 indicates that there are some objects that have to be transferred to other nodes. Hence, the synchronization cannot be successful since the transfer phase has to be repeated. This is the case shown in Figure 10, where an object is temporarily transferred from cube 1 to cube 2 before being definitely transferred to cube 3, since cube 1 does not have any reference to cube 3, being cube 1 and cube 3 not adjacent. The other phases illustrated in Figure 9 (movement, collision detection and management), already illustrated in Section 3.1, can be handled by each grid node without synchronization.

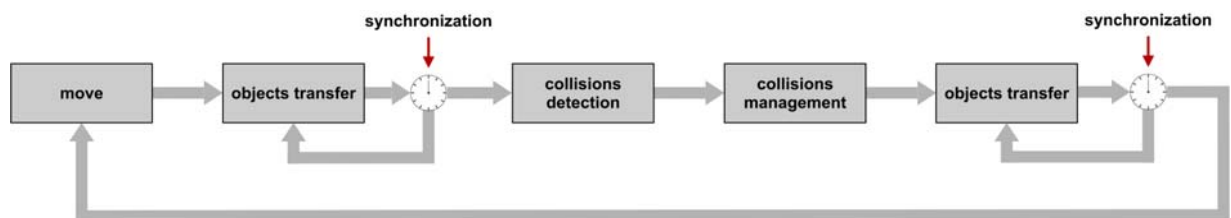

Figure 9 - The synchronization phases and their execution times.

The algorithm illustrated above guarantees the synchronization during the two phases of the object transfer. Since the synchronization routines unavoidably slow down simulations, when a distributed version of BiNS2 is executed, a significant performance degradation could be observed. As the number of grid nodes increases, the time spent in the synchronization phase increases as well, with a further performance degradation. For this reason, a suitable usage of the distributed version of BiNS2 is in very large simulations (when tens or hundreds of thousands objects have to be instantiated and simulated). In fact, in very large simulations, the performance degradation due to the synchronization of grid nodes is compensated by the advantage of managing, in each grid node, a smaller number of nano objects, thus benefiting of an increased level of parallelism. This is also consistent with the result of (1). The performance evaluation presented in section 4.2.2 shows a significant improvement when the number of objects is large, confirming these expectations.

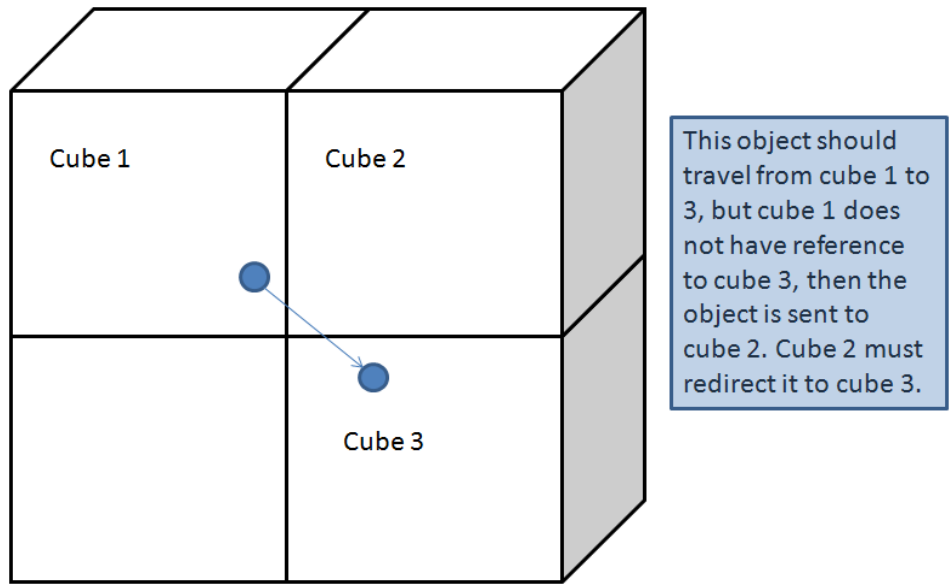

Figure 10 - An object moving to a destination grid node not adjacent to the initial grid node through an intermediate node during the same time step.

# 4 Use Case Analysis: the Interaction between Platelets and Endothelium via sCD40L

In this section, first we illustrate how to set up the simulation for modeling the communication mechanisms inside a blood vessel, by detailing the arrangement of endothelial cells on the vessel wall, the receptors placement on their surface, and the generation of blood cells. Then, we show the simulation results of the communication between platelets and endothelial cells via transmission of sCD40L carrier. Finally, we present an analysis of the performance of the grid implementation of our simulator.

## 4.1 Simulation Setup for Molecular Communications in Bloodstream

This section illustrates how the simulation for the study of atherogenesis has been set up, providing details about the placement of endothelial cells and relevant receptors, and how moving blood cells have been inserted into the simulation volume. The current version of the simulator is available at http://conan.diei.unipg.it/lab/index.php/research/17-nanonetworks.

### 4.1.1 Placement of the Endothelial Cells

The considered simulation scenario takes into account the propagation of the blood cells inside a blood vessel. A particular kind of these cells, the platelets, behave as transmitters, and the endothelial cells, that cover the inner vessel walls, are receivers. The surface of these cells facing the interior of the blood vessel is covered by several type of receptors. However, since we have simulated a single type of carrier, the sCD40L cytokine, we have instantiated only receptors compliant with it.

The endothelial cells have a regular flat shape [14] and are represented as cubic domains. Note that this is not an oversimplification, since this kind of cells that can be approximated by the square shape, which is the top view of a cube. The generation process of each cube creates a cube that is replicated and disposed along a thin layer of the vessel, by rotating it by $2\pi/N_h$ radians along the longitudinal axis of the vessel, where $N_h$ is the number of endothelial cells forming the border of a transversal section of the vessel, which is shown Figure 11. Finally, this layer is replicated along the longitudinal axis of the cylinder which models the vessel in order to model the whole vessel inner surface.

The size of each cube approximates the size of an endothelial cell, which is determined so as to dispose a integer number of cubes along the inner border circumference of the vessel. This is a satisfactory model, since although real biological cells may be different from each other, their width is in the order of 10-20 μm [14]. In addition, it results from the experiment described in [37], whose details are reported in [5], that the side of used endothelial cells is in average equal to 15 μm.

In the considered case, we have selected a vessel radius $R$ equal to 30 μm, which is compliant with the size of a venule, one of the thinnest human blood vessels, whereas the size of the endothelial cells, $d_h$, has been set equal to 15 μm. The length of the border circumference of the vessel is $c = 2\pi R = 188,5 \mu m$. The ratio $c/d_h = 12,57$, hence we have selected $N_h = 13$ endothelial cells to approximate the circumference $c$. The resulting approximated width of the endothelial cells is $\tilde{d}_h = c/N_h = 14,5 \mu m$. This width of the endothelial cells is thus compliant with the value obtained from the experiments described in [5][37].

The distance $V_h$ of the center of the cubes ($C_h$) from the vessel axis is shown in Figure 11. Most of the volume of the cube is outside the vessel, and it does not affect the simulation, since we focus only on the inner volume of the vessel. In the simulations, nano objects can collide with the vessel wall, but not with the cube sides. The following equations can be easily derived by observing Figure 11 and using geometric considerations (to simplify the picture, we have shown the case for $N_h=6$):

$$apothem = R\cos\left(\frac{\pi}{N_h}\right) \cong 29{,}13\,\mu m \tag{20}$$

$$V_h = apothem + \frac{\tilde{d}_h}{2} \cong 36{,}38\,\mu m. \tag{21}$$

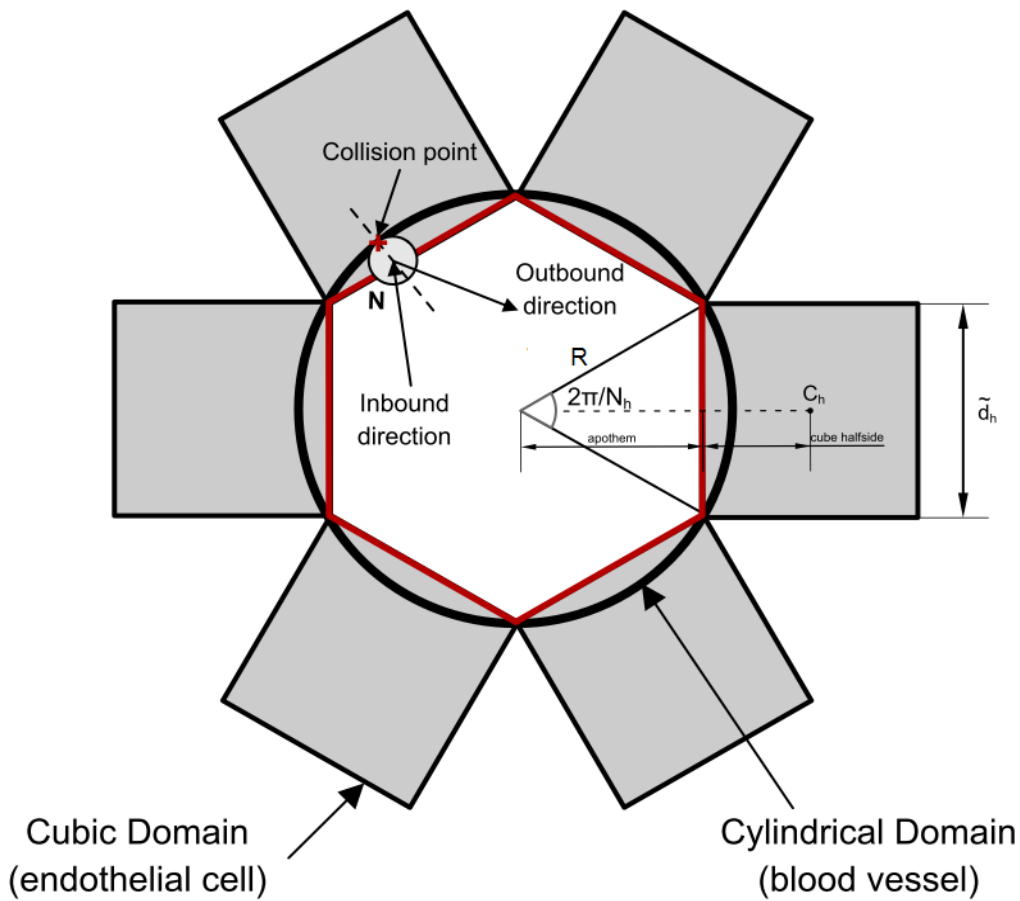

Figure 11 - Placement of the endothelial cells as cubic domains, and collision of nano-objects with the vessel walls.

It is important to clarify that the simulated endothelial cells reside *on* the cylinder surface, and not on the cube surface. The side of the cube is a virtual surface, which does not cause any bounce. The rationale for using the cube is that, although the modeled endothelial cell resides on the cylinder side surface, in this way the collision of a nano object with that cell can occur *only* if that object is inside the cube domain relevant to that endothelial cell. Thus, the usage of cube domains simplify the management of the collisions of nano objects with the endothelium, including also an easier handling of the assimilation process, since the receptors associated to an endothelial cells are *only* those inside the cube. The next Section 4.1.2 illustrates how receptors are placed on the surface of the cylinder.

### 4.1.2 Placement of the Receptors

The surface of each endothelial cell facing the interior of the vessel holds 1000 receptors [23], which are randomly disposed over it. Since the endothelial cells in real blood vessels have a curved surface that follows the circular shape of the vessel, we have projected each receptor on the vessel surface, even if it still belong to its cubic domain modeling the endothelial cell (the vessel surface where the receptors are projected is inside the cube). For each cube modeling an endothelial cell, we consider the secant plane that passes through the cube center ($C_h$) and is orthogonal to the longitudinal axis ($\hat{d}$) of the cylinder modeling the vessel, shown in Figure 12. This plane is clearly orthogonal also to the cube surface that exposes the receptors. Each receptor has a distance to the center of the cube equal to $\vec{V}_{R1}$ (see again Figure 12, side view). By observing the side view of Figure 12, it is easy to verify that

$$\vec{V}_{R2} = \vec{V}_h + \vec{V}_{R1}. \tag{22}$$

We define γ as the complementary angle of $\beta$ formed by $\vec{V}_{R2}$ and the longitudinal axis $\hat{d}$, and $\vec{V}_{R2\_NEW}$ as the projection of the radius of the cylinder $\vec{R}$ that lies on the secant plane on $\vec{V}_{R2}$. This new vector is parallel to $\vec{V}_{R2}$ and passes through the center of the receptor, but terminates on the cylinder surface:

$$\left|\vec{V}_{R2\_NEW}\right| = \frac{\left|\vec{R}\right|}{cos(\alpha)}. \tag{23}$$

Expressing this new vector as a function of the center of the cube $C_h$, we obtain $\vec{V}_{R1\_NEW}$, that gives the projection of that receptor on the cylinder surface:

$$\vec{V}_{R1\_NEW} = \vec{V}_{R2\_NEW} - \vec{V}_h. \tag{24}$$

If the projection phase places any receptors outside the cube, then those receptors are removed, as shown in Figure 13. This check is done by evaluating the vector $\vec{P}$ given by:

$$\vec{P} = \vec{V}_{R2\_NEW}\, sin(\gamma). \tag{25}$$

If $-d_h/2 < |P| < d_h/2$ the considered receptor can be placed on the new coordinates, otherwise it is definitely removed.

The collisions between every nano object (blood cells and cytokines) with the endothelial cells are handled in the same way, by evaluating the collisions only when they happen with the cylindrical surface and not with the cube surface. This procedure is implemented by delegating the handling of the collision between a nano object and a cubic domain to its parent cylindrical domain (the vessel). In this way, the detection of the point of impact and the possible bounce is evaluated on the correct position on the cylindrical domain surface, as shown in Figure 11. If that collision causes an assimilation by a specific receptor of the cube domain located on the impact point, the bounce will not be evaluated and the receiving procedure will be started. Clearly, this can happen only if the nano object is a carrier.

The other cell types that flow along the vessel are platelets, white, and red blood cells. They affect the simulation only in the propagation phase, colliding with each other and with the other cell types.

Finally, platelets can emit bursts of carriers, in order to model the following phenomenon. When an external stimulus is received by platelets, they activate and express CD40L on the cell surface. As mentioned above, the expressed CD40L is subsequently cleaved and shed from the platelet surface as sCD40L. The size of the burst has been determined by analyzing the results of a set of laboratory experiments, as explained in Section 4.2.1.

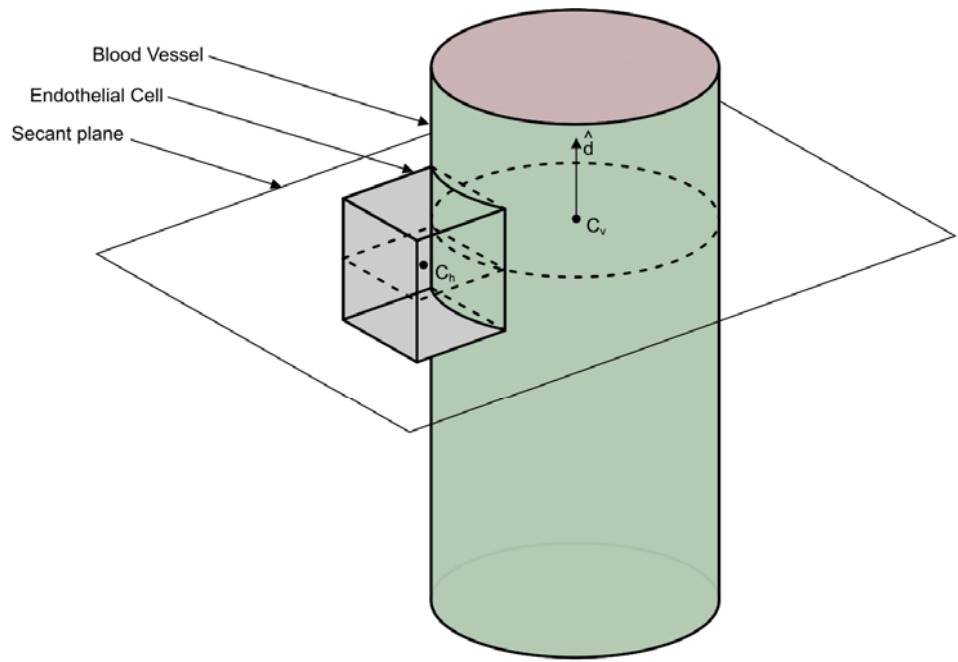
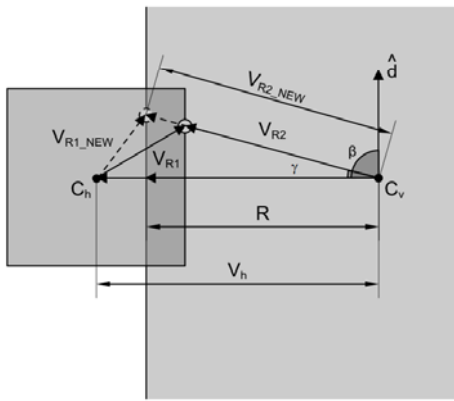
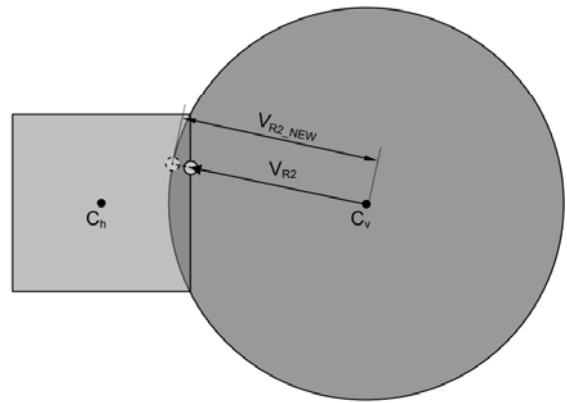

Side view

Top view

Figure 12 - Projection of the receptors on the curved surface of the vessel.

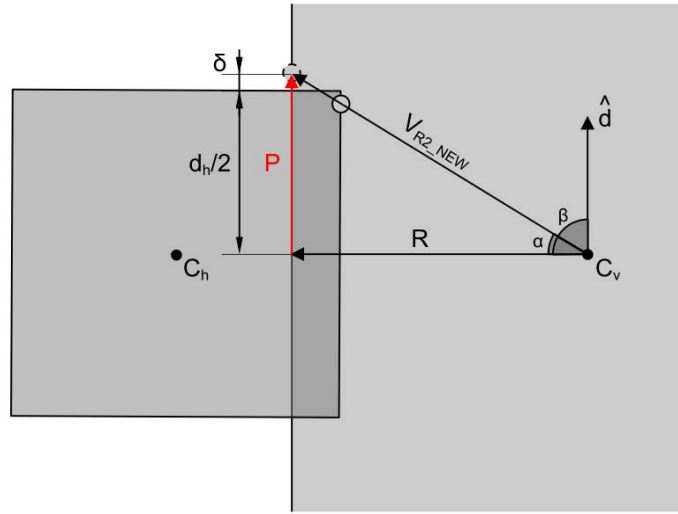

**Side view**

Figure 13 - Removal of a receptor projected outside the cube domain.

### 4.1.3 Positioning of Blood Cells

The case study that has been analyzed needs the definition of a special type of cylindrical domain, in order to accurately represent the behavior of the blood vessel. We have disables the collision checks on the top and bottom surface of the cylinder and destroys any nano object that leaves the considered section through these surfaces. The collisions with the side surface are handled as illustrated in paragraph 3.2.2.

We have defined a thin volume at the beginning of the vessel where the nano objects are created. This thin volume is randomly filled by a deterministic amount of blood cells in order to generate a realistic concentration (we have considered platelets, white blood cells, and red blood cells). Then, this volume is replicated, shifted, and rotated many times to fill up the blood vessel, as shown in Figure 14. This procedure speeds up the simulation transient phase, since generates a blood vessel filled of cells since the beginning of the simulation. After a small transient period, set equal to 40000 time steps, having verified that the nano objects spatial distribution assume the parabolic profile given by the Poiseuille flow theory [25][26][27][28], we place the transmitter node, which acts as a platelet, and releases a single burst of carriers (sCD40L) [37].

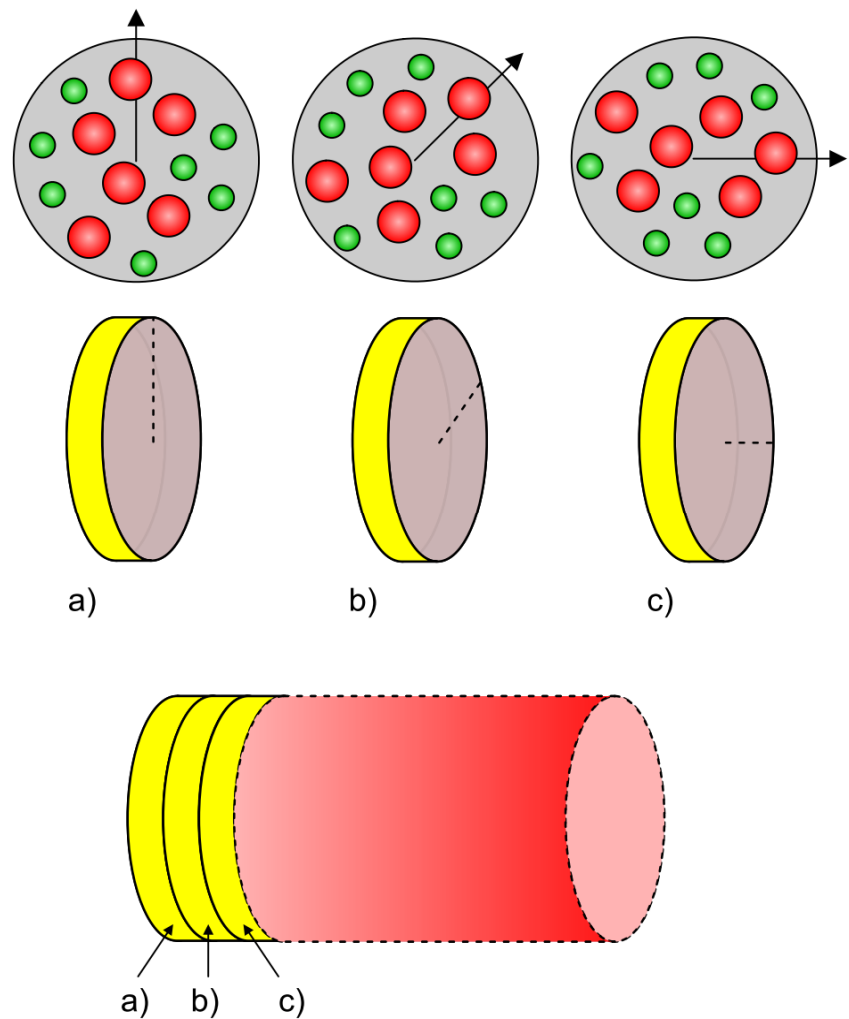

Figure 14 - The strategy used to initialize the simulation, by filling the blood vessel with blood cells.

In order to have the control on the transmission coordinates, we create the transmitter node at a fixed position[1], it suddenly emits, and then it is free of move according to the blood flow. The selected set of pre-established transmitter positions are at 400 μm from the first endothelial cells along the blood vessel, as shown in Figure 15. Periodically, new nano objects are created at the beginning of the vessel, in the region labeled as "continuous creation" in Figure 15. In this way the average concentration of blood cells is maintained nearly constant in each phase of the simulation, with a small variance, in order to balance the destruction of the cells exiting the simulated section of the vessel. Since we are simulating venules, which are located at the peripheral of the cardio circulatory system, the effect of the heartbeat is negligible, and we can approximate the flow pressure as constant. This creation process is relevant only to blood cells, since in this work we consider the emission of a single burst of carriers at 40000 time steps.

---

[1] If the selected position is (partially) occupied by one or more other objects, this attempt is repeated until it is successful.

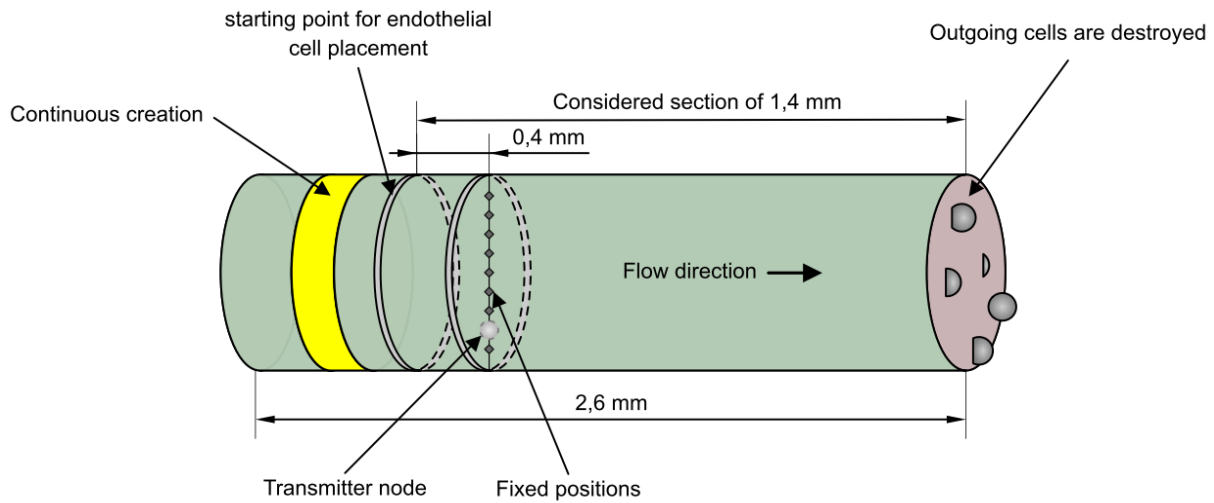

Figure 15 – Sections and points of interests of the simulated blood vessel.

As for the behavior of nano objects in the simulated blood flow, it is known that the massive presence of red blood cells creates a so-called Casson profile [26], whereas the carriers propagations can be modeled as the sum of Brownian and convection contributions, producing an *effective* Brownian diffusion motion with a larger contribution along the flow propagation direction [25][26][27]. However, as confirmed by laboratory experiments, the recent paper [28], which models the interactions between carriers and blood cells, shows through computer simulations that carriers are pushed towards the vessel walls by the stream of red blood cells. We follow the approach proposed by [28] by explicitly taking into account inelastic collisions and bounces between carriers and blood cells. This means that the cells in the blood flow strongly modifies the propagation of carries within the blood vessel.

The parameters used in the simulations are reported in Table 1.

## 4.2 Numerical Results

### 4.2.1 Numerical Results for the Communications in Blood Vessels

In this Section, we present the numerical results relevant to the simulation of the communications between the mobile transmitter node and fixed receivers positioned on the vessel walls. The communication is characterized in terms of both the footprint of the transmitted signal on the receivers set and the time needed to trigger the decoding of the information signal at the receivers, as a function of the receivers sensitivity. The transmitted signal models an impulse, since it consists of a burst $B$ of carriers, modeling the release of sCD40L.

Table 1 - Simulation parameters.

| Vessel parameters | | Endothelial cells | |
|---|---|---|---|
| Length | 2.6 mm | Cell side | 15 μm |
| Radius ($R$) | 30 μm | Number of receptors | 1000 |
| Mean flow velocity | 0.5 mm/s | Receptors radius | 4 nm |
| Viscosity | 0.0013 Pa × s | **White blood cells** | |
| Temperature | 310° K | Concentration | $4 \times 10^3$ mm$^{-3}$ |
| **Platelets** | | Receptors | 1000 |
| Concentration | $2 \times 10^5$ mm$^{-3}$ | Radius | 5 μm |
| Radius ($r_p$) | 1 μm | **Red blood cells** | |
| Number of receptors | 1000 | Concentration | $4 \times 10^6$ mm$^{-3}$ |
| Receptors radius | 4 nm | Radius | 3.5 μm |
| Burst size | 3000 | **Simulation parameter** | |
| sCD40L radius | 1.75 nm | Time step | 5 μs |

Figure 16 shows the total number of endothelial cells able to decode the signal transmitted by the mobile transmitter, as a function of the time elapsed since the transmission of the carriers. We have characterized the receiver by a threshold $S$, which is the sensitivity to the received signal: it is the minimum number of carriers received by an endothelial cell necessary to decode the signal. We have used four $S$ values: 1, 2, 5, and 10. In addition, each sub-figure is relevant to a different transversal position of the transmitter: $L0$ indicates a transmitter close to the vessel wall, $L5$ a transmitter positioned at the center of the vessel section. By analyzing the results of the experiment described in [5], we have estimated that the average number of sCD40L carriers emitted by a stimulated platelet is 3067. Thus, the value used in our simulations (3000) is realistic.

The first observation is that, in the considered section of the blood vessel, the number endothelial cells able to activate, i.e. to decode the signal, increases with the time after the signal transmission, with a nearly linear pattern. In addition, for any positions of the transmitter closer to the walls, the slope of the curve is higher. These results are rather expected as when carriers are released close to the wall, due to the effect of their Brownian motion combined with the collisions with the flow of larger blood cells (which is the dominant contribution), they are spread around, significantly towards the vessels wall, where the fixed receivers reside (see also the results in [28]). As the position of the transmitter is closer the center of the vessels, the released carriers are still spread around but, since the vessel wall is more distant, some of them are trapped by multiple rebounds in the flow of blood cells and are taken away by the flow itself. Thus, a lower number of endothelial cells receives a signal intensity able to trigger the signal decoding.

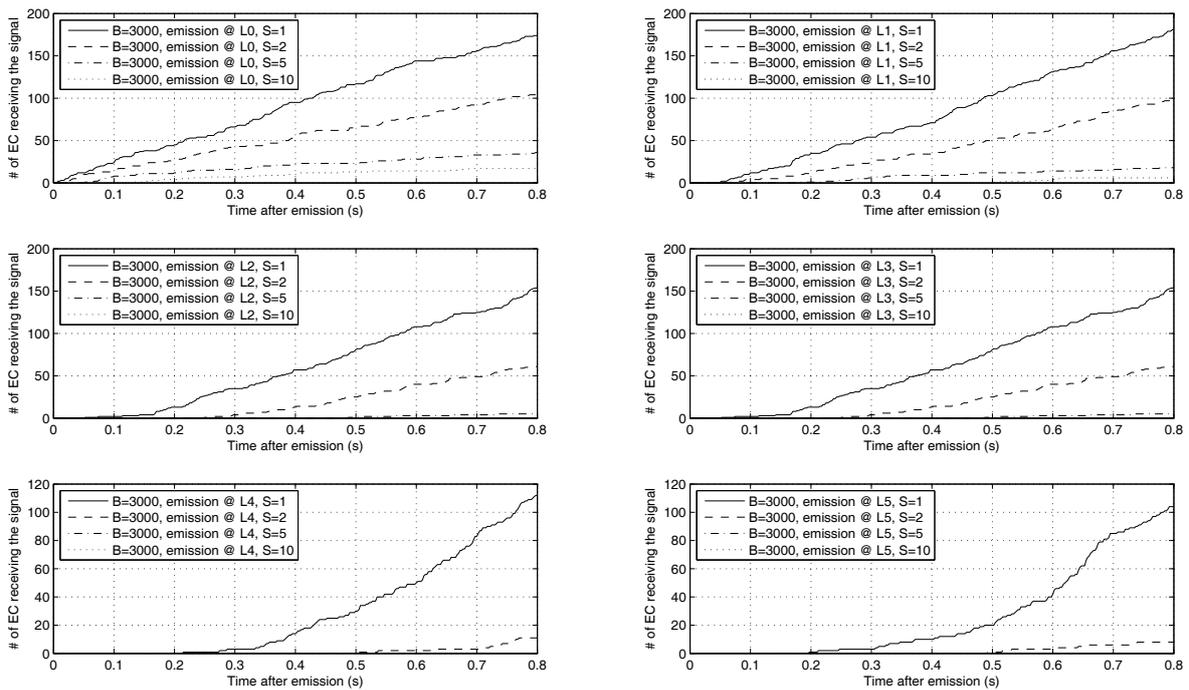

Figure 16 – Number of receivers decoding the transmitted signal as a function of the time, for a burst size B=3000 carriers. The parameters are the decoding threshold *S*, the transmission position (*L0 – L5*).

In order to gain more insight about the effects of the communication channel, we have analyzed not only the number of endothelial cells able to decode the signal, but also their displacement on the vessels wall with respect to the transmitter position, and the time of activation since the release of the burst, as a function of the receiver position. For this purpose, we identify a receiver by its cylindrical coordinates, given by the angle $\phi$ and the longitudinal coordinate $E_z$, as shown in Figure 17. Note that since all receivers are disposed at a constant distance from the $z$ axis on the cylinder surface, it is not necessary to specify the radial coordinate $r$, which has a constant value equal to $R$. As for the transmitter, the cylindrical coordinates of its center are ($\phi=0$, $r$, $z=0$), where $r$ values belong to the set from $L5$ ($r=0$) to $L0$ ($r=R-r_p$, see Table 1).

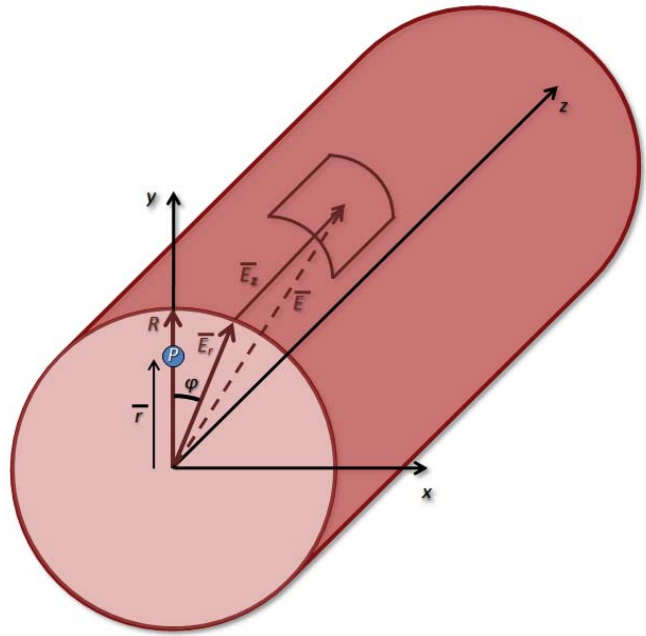

Figure 17 – Endothelial cell identified by means of the cylindrical coordinates of its center ($\phi$, $r=R$, $z=E_z$).

The results of this analysis are shown in Figure 18, where the cylindrical coordinates of the receivers are expressed in radians for the angle $\phi$ and in µm for the longitudinal coordinate $z$, whereas the chromatic bar indicates activation times in seconds. This figure could be regarded as a sort of radiation pattern with the additional information about activation time. This figure is relevant to the case with $B=3000$ and $S=2$. The transmitter position is relevant to the most significant cases, that is the one closest to the wall ($L0$), an intermediate position ($L3$), and the position at the center of the vessel ($L5$). Since the number of endothelial cells for a given longitudinal coordinate is odd ($N_h$=13, see (23)), there are small asymmetries with respect to the $\phi$ coordinate in the patterns shown in Figure 18. As the reader can observe, the closer the transmitter to the vessel walls, the lower the time to activate surrounding receivers. This information can be deduced also by analyzing Figure 16. However, Figure 18 provides additional significant information. First of all, it shows which endothelial cells are the first to decode the signal. Let us consider Figure 18.a, where the transmitter is in position *L0*. As expected, the cells that can first decode the signal are those closest to the transmitter. They are identified by an angle $\phi$ and a longitudinal coordinate *z* close to 0. Surprisingly, although carriers are dragged by the blood flow, the collision effects with blood cells, mainly red blood cells, is so strong that even some endothelial cells which are behind the transmitter (i.e. with a negative longitudinal coordinate) can be activated. In addition, the proximity to the vessel wall does not allow carriers to propagate in the direction of the $\phi$ coordinate, and the radiation pattern results narrow and long. In more detail, the footprint is well contained within the range [-1,1] radians and [-400,650] μm. As for the activation time, it is clear that the larger the distance from the transmitter, the larger the time to activate the receivers, which, however, is within 1 second. The first activated cells (by approximately 25 μs) have a symmetrical displacement around the transmitter position. As for the cells activated later, the exhibited pattern extends more deeply towards the positive values of the longitudinal coordinate, due to the push exerted by the blood flow. Now, let us focus on the footprint relevant to the transmitter position *L3* (Figure 18.b). Two effects are evident. The first is that the footprint is broader both on the longitudinal axis and (especially) on the $\phi$ angle, being contained in the range [-2, 2] radians and [-350, 850] μm. Thus, the footprint is shifted towards positive values of z. The second is that, even if the patterns it is still nearly symmetric with respect to the z axis, the shape has changed, with a larger section for positive values of the z coordinate. Both these phenomena are due to an increased distance of the release point of the carriers from the vessel walls, which increases the effects of both the flow dragging towards positives z values and the spreading due to the collision with red blood cells. As for the activation time, it exhibits the largest variability, ranging from 0.22 to about 2.36 seconds. The third subfigure (Figure 18.c) is still different. In this case, the emission point is located on the vessel axis *z*. This leads to a (nearly) symmetrical and broad profile of activation on the vessel walls.

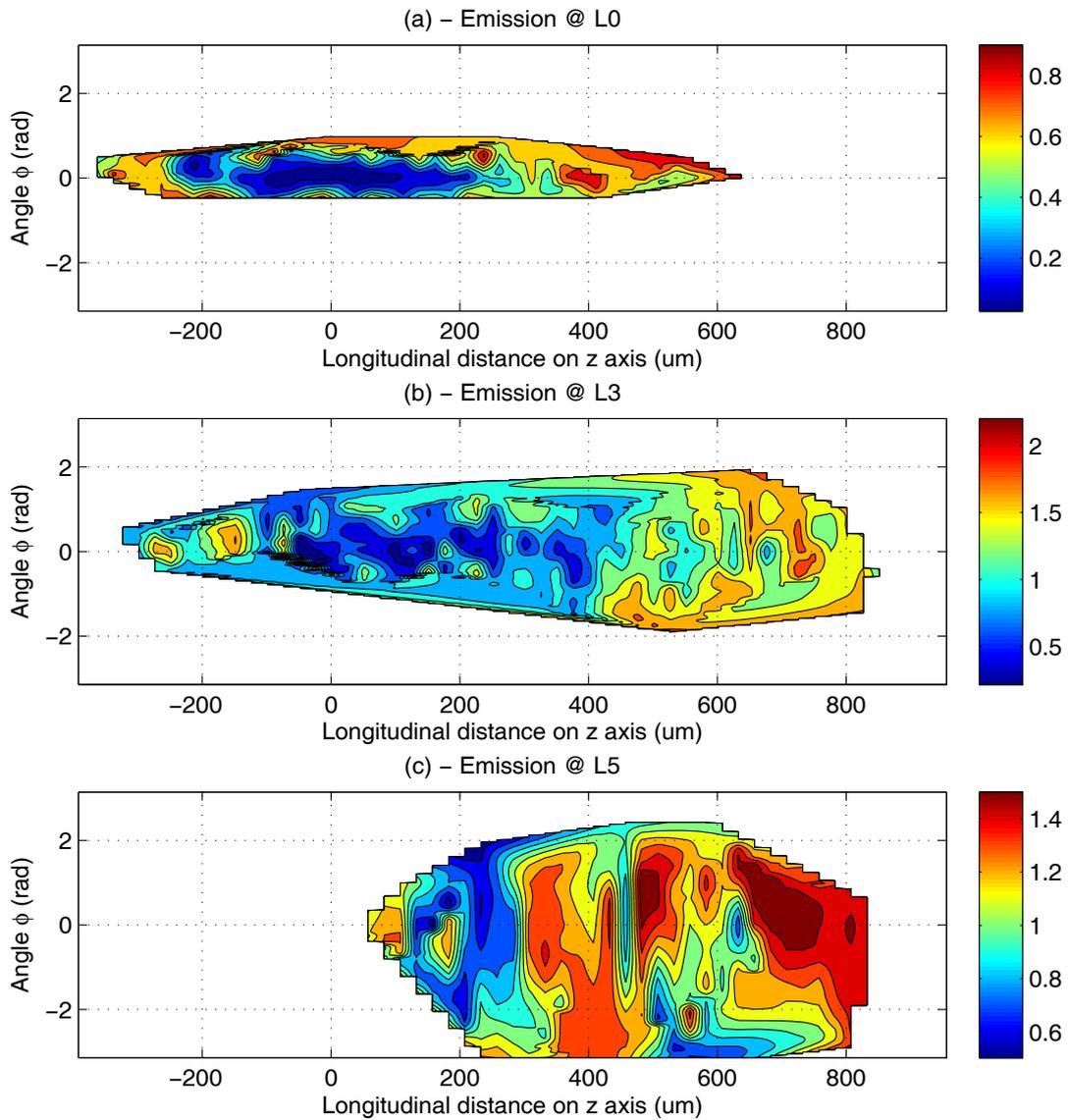

Figure 18 – Activation pattern and time of activation of fixed receivers as a function of their cylindrical coordinates ($\phi, E_z$), for different positions of the transmitter.

Only cells with a positive z coordinate are activated, and the variability ranges are about [-π, π] radians for the angle $\phi$ and [50, 950] μm for z, thus covering all the span of $\phi$. The first activation time is clearly larger than the values previously observed (approximately 0.5 seconds), but the relevant variability has decreased in comparison with the one observed when the transmitter was located in the position *L3*, since last cells activate at approximately 1.6 seconds. This behavior can be explained. Being the transmitter located on the lungitudinal axis of the blood flow, the flow speed is maximum. In addition, the transmitter point is equidistant from all vessels wall, independently by $\phi$. Making a parallel with radio communications with a directive antenna with a radiation pattern pointing towards positive values of z, the higher the antenna height (i.e. the value of the *r* coordinate of the transmitter), the larger the distances covered by the signal and thus the broader the signal footprint in terms of $\phi$ and *z*.

### 4.2.2 Numerical Results for the Grid Implementation of BiNS

In order to evaluate the difference between the grid deployment of the simulator and the classic approach, we have done several simulations of a bounded cylindrical domain. Simulations have been performed by using two servers, each equipped with 4 Six-Core AMD Opteron™ processors 8425HE@2.1GHz (24 CPU cores) and 64GB of RAM, connected through a Gigabit Ethernet switch. The operating system is the linux Ubuntu server 12.04, 64-bit version, and we used Oracle Java 7 and GridGain 4. Each simulation is configured to use 64 threads. For the grid approach, we have considered two cases. The first is used to evaluate the performance of several grid nodes deployed on a single server, in order to fully benefit of the multicore architecture, which provides many CPU cores. The second is used to evaluate the performance of several grid nodes equally distributed over two identical servers. Clearly, this is a special case of computing grid, acting more as a computer cluster than a grid. However, we have performed successful tests using computers in different locations, interconnected by a geographical network, as usually happens with standard computational grids. In this latter case, the main benefit of using the grid approach is the possibility of increasing the scale of the simulation.

The simulations on a single server have been performed by using two configurations, which differ only by the number of objects in the medium: about 10000 in one case and 50000 in the other. The simulation environment is a cylinder with a height of 72 μm and a radius of 393.3 μm. Mobile objects are carriers only. The cylindrical domain is split by 2 to 16 smaller cubic domains, which lie on a single layer (Figure 19). In case the grid is deployed by using two servers, and the splitting algorithm generates up to 36 subdomains per layer. We have collected the results of the steady state in order not to include in the comparison the transient due to the initialization phase, which is time consuming since it executes the splitting process. Figure 20 shows the time needed to accomplish 1 second of simulation as a function of the number of cubic subdomains used on a single layer. Note that the ordinate axis, which reports hours of computation, uses a logarithmic scale. As expected, the benefits provided by the grid approach are more appreciable when the number of objects is larger and many grid nodes are used. In fact, having a large number of objects to manage requires long computational time to handle movements and collisions, thus the gain achievable by parallelizing the computation allows compensating the grid overhead due to synchronization. Since increasing the number of grid nodes on each layer decreases significantly the number of objects to be managed by each subdomain, increasing the number of grid nodes decreases the computational time (see equation (1)). This is true until the number of grid objects in each subdomain is sufficiently large. When it is too small, the time saved by parallel execution becomes comparable with the time overhead due to synchronization, and any further increase of the grid node number causes a performance degradation.

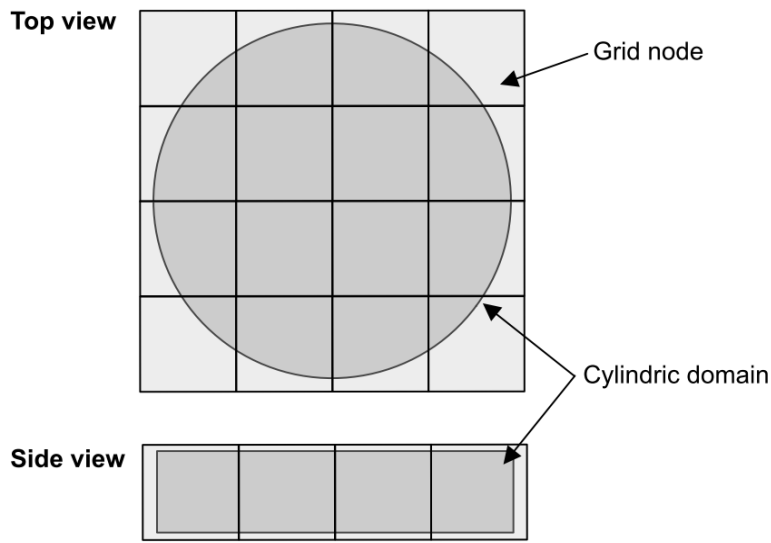

Figure 19 – Splitting of the cylindrical domain in grid simulations, top and side view.

This is particularly evident in the first case shown in Figure 20, relevant to a simulation with about 10000 objects and a single server. Note the 10000 objects is not an extremely high number also for a single node, thus when only two grid nodes are used on a single layer, each of them has still to manage a large (but not excessive) number of objects. Hence, the gain achievable by parallelizing the execution is comparable with the relevant overhead, which slightly dominates. As the number of subdomains increases, the parallelization becomes effective up to 8 grid nodes per layer. Beyond this value, the overhead becomes dominant again. In addition, we have seen that deploying more than 16 cubes per layer is not feasible on a single server, since the system crashes. The second case is relevant to a simulation including about 50000 objects running on a single server. In this case, due to the larger number of objects, the grid is effective also with 2 cubic subdomains per layer, and continues to be effective up to the maximum number of objects per layer, that is 16. It exhibits a computational time which is up to 16 times shorter than the one achieved by using a single instance. Finally, in order to go beyond 16 cubic domains per layer, we have deployed the grid over two identical servers. In this case, the overhead due to the grid synchronization, but even due to the transfer of grid objects between grid instances, is larger, since the communication is not confined within the server but involves the network. We can see that for a small number of cubic subdomains, the implementation with 2 servers is affected by this additional overhead and performs worse than in the single server case. However, when the number of grid node increases (starting from 8 cubic subdomains per layer), balancing the load on two servers becomes (slightly) effective and this leads to a small improvement. Since any further splitting increases the overhead, the performance slightly degrades for abscissa values larger than 16.

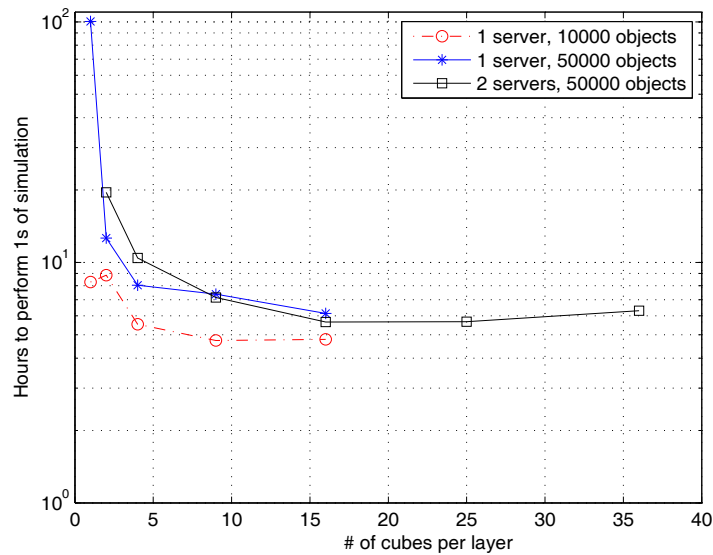

Figure 20 – Execution time for grid simulation as a function of the number of subdomains per layer.

In this regards, we have measured the traffic exchanged between the two servers hosting the grid instances during simulation execution in steady state conditions, as a function of the number of grid instances. As expected, the higher the number of nodes, the larger the network overhead. The overhead increases more than linearly since, when many nodes are randomly distributed over different servers, the probability that a communication between any two grid nodes is confined within a single server decreases. This also implies that, for larger simulations, the grid is effective only if it is implemented through a high speed LAN, otherwise the network behaves as a bottleneck slowing the simulation down. Also, running grid nodes managing adjacent cubes in the same server could limit the network overhead.

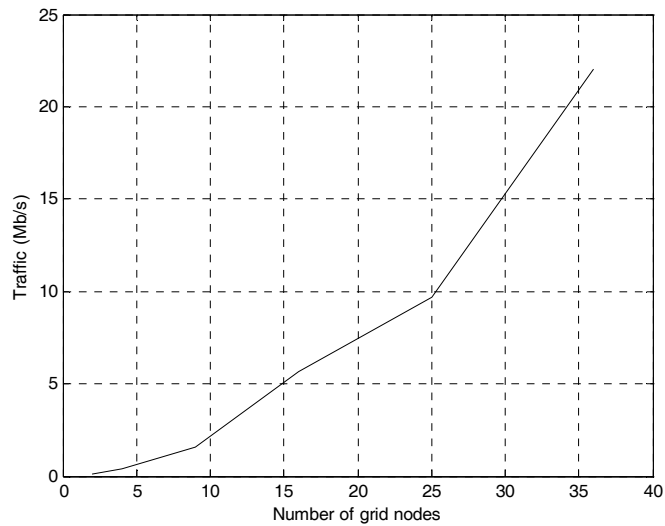

Figure 21 – Total traffic for grid simulation as a function of the number of grid nodes.

Although these results on the grid deployment are preliminary and further analysis will be made on more complex scenarios, a number of considerations apply. First, since deploying multiple grid nodes on a single server allows decreasing the computation time, it is evident that, notwithstanding the improvements done with respect to the previous version of BiNS in terms of parallelism [3], the simulation has still a number of phases where the computation is serial, which slow down the execution. We think that we can cope with this issue by deploying multiple cubic subdomains within each grid node, by adopting an octtree structure to strongly parallelize the computation [38], possibly re-adapting the splitting during the simulation execution. Clearly, this would allow to run a grid node per server, thus effectively exploiting the available computing resources. The second action line consists of improving the bottleneck of the grid platform. In the current form, for handling the synchronization phases we have used a module already present in the GridGain package [40]. Given the peculiarities of our framework we think that it is possible to optimize the synchronization phase. It can be done by defining an ad hoc, simple application protocol, in which the client side runs on the grid nodes, and the server part, controlling the simulation progress, is be executed either on a specific grid node or on an external, dedicated node. A possible hosting environment for this synchronization architecture could be the JSLEE (Java APIs for Integrated Networks Service Logic Execution Environment, [58]) applications server, which is characterized by high throughput and low latency. Although the JSLEE specifications have been defined for creating platforms able to run carrier-grade telecommunications services, such as VoIP calls management ones (see e.g. [44]), it can also be used to manage any type of data or signaling (see e.g. [45]), and it comes with powerful tools able to speed up software programming [46]. A possible alternative solution to using a central JSLEE server or the current grid implementation could be the adoption of epidemic protocols to transport signaling [43], if a decentralized architecture is preferred and/or needed.

Finally, a further direction for improving performance is to use not only standard CPUs of multicore servers but also graphical processing units (GPUs) [41]. Please note that this solution is becoming more and more common in supercomputers, since it is able to speed up performance significantly [42].

# 5  Conclusion and Future Work

In this paper, we have presented an upgraded version of the BiNS simulator, able to model nano-scale communications based on Brownian diffusion with drift in a bounded space. The simulator is able to represent partially inelastic collisions between biological entities of different size. We have also enhanced the simulator in order to model the basic communications between mobile transmitters and fixed receivers in blood vessels, which is a suitable model of carrier exchange between platelets and endothelium, a mechanism of paramount importance in the initial stages of atherosclerosis.

Simulation results show the effects of both the signal intensity and the transmitter position on the number and the locations of receivers able to decode the transmitted signal. Also, the distance between the transmitter and the surface where receivers are deployed has a strong effect, in a way similar to the transmission of radio signals by directive antennas. The directionality of the transmission is not provided by the transmitter through directive antennas, but is generated by the effects of the Brownian motion, blood flow force, and collisions of information carriers with blood cells.

Finally, we have presented a grid implementation of the simulator, able to significantly reduce the simulation time in particular system configurations, by also highlighting future development directions in order to further decrease the computational time.

Future works will focus on the use of the simulator to implement suitable mathematical models for both channel and receivers. Due to the presence of blood cells, the results shown in this paper significantly deviate from the models previously proposed in the literature. This research will be assisted by a number of biological experiment, in order to accurately calibrate the simulator. In addition, we will analyze also the spreading properties of the channel, in order to design suitable transmission patters and receivers able to coexist by using a shared channel.

## References

[1] I.F. Akyildiz, F. Brunetti, C. Blázquez, "Nanonetworks: A new communication paradigm", Computer Networks, Elsevier, 52(12), August 2008.

[2] B. Atakan, O.B. Akan, S. Balasubramaniam, "Body area nanonetworks with molecular communications in nanomedicine", IEEE Communications Magazine, January 2012.

[3] L. Felicetti, M. Femminella, G. Reali, "A simulation tool for nanoscale biological networks", Nano Communication Networks, vol. 3, no. 1, 2012, pp. 2–18, DOI: 10.1016/j.nancom.2011.09.002.

[4] I.F. Akyildiz, J.M. Jornet, "The Internet of Nano-Things", IEEE Wireless Communication Magazine, 17(6), December 2010.

[5] L. Felicetti, M. Femminella, G. Reali, P. Gresele, and M. Malvestiti, "Experimental campaign on the in-vitro platelet-endothelial cells interactions", Tech. Rep. [Online]. Available: http://conan.diei.unipg.it/pub/experiment.pdf.

[6] S. Kadloor, R. Adve, A. Eckford, "Molecular communication using brownian motion with drift", IEEE Transactions on NanoBioscience, vol. 11, no. 2, pp. 89–99, June 2012.

[7] M. Pierobon and I. F. Akyildiz, "A physical end-to-end model for molecular communication in nanonetworks", IEEE Journal on Selected Areas in Communications, vol. 28, no. 4, pp. 602–611, May 2010.

[8] I. F. Akyildiz, J. M. Jornet, and M. Pierobon, "Nanonetworks: a new frontier in communications", Communications of the ACM, vol. 54, no. 11, pp. 84–89, Nov. 2011.

[9] E. Gul, B. Atakan, and O. B. Akan, "NanoNS: a nanoscale network simulator framework for molecular communications", Nano Communication Networks, vol. 1, no. 2, pp. 138–156, 2010.

[10] N. Garralda, I. Llatser, A. Cabellos-Aparicio, E. Alarcon, M. Pierobon, "Diffusion-based physical channel identification in molecular nanonetworks", Nano Communication Networks, vol. 2, no. 4, pp. 196–204, Dec. 2011.

[11] D. Gidaspow, J. Huang, "Kinetic Theory Based Model for Blood Flow and its Viscosity", Annals of Biomedical Engineering, vol. 37, no. 8, August 2009, DOI: 10.1007/s10439-009-9720-3

[12] A.S. Leroyer, P.-E. Rautou, J.-S. Silvestre, Y. Castier, G. Lesèche, C. Devue, M. Duriez, R.P. Brandes, E. Lutgens, A. Tedgui, C.M. Boulanger, "CD40 Ligand+ Microparticles From Human Atherosclerotic Plaques Stimulate Endothelial Proliferation and Angiogenesis - A Potential Mechanism for Intraplaque Neovascularization", Journal of the American College of Cardiology, 2008, 52(16), pp.1302-1311, DOI:10.1016/j.jacc.2008.07.032.

[13] T. Nakano, M. Moore, F. Wei, A. Vasilakos, and J. Shuai, "Molecular communication and networking: Opportunities and challenges," IEEE Transactions on NanoBioscience, vol. 11, no. 2, pp. 135 –148, June 2012.

[14] R. A. Freitas, Nanomedicine, Volume I: Basic Capabilities. Landes Bioscience, Georgetown, TX, 1999.

[15] D.A. Reasor Jr., M. Mehrabadi, D.N. Ku, C.K. Aidun, "Determination of Critical Parameters in Platelet Margination", Annals of Biomedical Engineering, 2012, DOI: 10.1007/s10439-012-0648-7

[16] S. Giannini, E. Falcinelli, L. Bury, G. Guglielmini, R. Rossi, S. Momi, and P. Gresele, "Interaction with damaged vessel wall in vivo in humans induces platelets to express CD40L resulting in endothelial activation. No effect of aspirin intake", Am J Physiol Heart Circ Physiol, vol. 300, no. 6, pp. H2072–9, 2011.

[17] Z. Kaplan and S. Jackson, "The role of platelets in atherothrombosis." Hematology Am Soc Hematol Educ Program, pp. 51–61, 2011.

[18] U. Schonbeck and P. Libby, "The CD40/CD154 receptor/ligand dyad," Cell. Mol. Life Sci., vol. 58, no. 1, pp. 4–43, January 2001.

[19] V. Henn, J. R. Slupsky, M. Grafe, I. Anagnostopoulos, R. Forster, G. Muller-Berghaus, and R. A. Kroczek, "CD40 ligand on activated platelets triggers an inflammatory reaction of endothelial cells," Nature, vol. 391, no. 6667, pp. 591–594, February 1998.

[20] Y. Chen, J. Chen, Y. Xiong, Q. Da, Y. Xu, X. Jiang, and H. Tang, "Internalization of CD40 regulates its signal transduction in vascular endothelial cells," Biochem Biophys Res Commun., vol. 345, no. 1, pp. 106–17, June 2006.

[21] P. Andre´, L. Nannizzi-Alaimo, S. K. Prasad, D. R. Phillips, "Platelet-derived CD40L: the switch-hitting player of cardiovascular disease", Circulation, vol. 106, no. 8, pp. 896–899, Aug. 2002.

[22] F. Mach, U. Schonbeck, and P. Libby, "CD40 signaling in vascular cells: a key role in atherosclerosis?" Atherosclerosis, vol. 137, Supplement 1, pp. S89 – S95, 1998.

[23] D.A. Lauffenburger, J.J. Linderman, "Receptors; models for binding, trafficking, and signaling", Oxford University Press, New York, 1993, ISBN 0-19-506466-6.

[24] J. Philibert, "One and a half century of diffusion: Fick, Einstein, before and beyond," Diffusion Fundamamentals, 4 (2006), pp. 6.1–6.19.

[25] G.Taylor, " Dispersion of Soluble Matter in Solvent Flowing Slowly through a Tube", Proc. R. Soc. Lond. A 1953 219, 186-203, DOI: 10.1098/rspa.1953.0139.

[26] P. Decuzzi, F. Causa, M. Ferrari, and P. Netti, "The effective dispersion of nanovectors within the tumor microvasculature," Annals of Biomedical Engineering, vol. 34, no. 4, pp. 633–641, Apr. 2006.

[27] F. Gentile, M. Ferrari, and P. Decuzzi, "The transport of nanoparticles in blood vessels: the effect of vessel permeability and blood rheology." Annals of Biomedical Engineering, vol. 36, no. 2, pp. 254–61, February 2008.

[28] J. Tan, A. Thomas, and Y. Liu, "Influence of red blood cells on nanoparticle targeted delivery in microcirculation," Soft Matter, vol. 8, pp. 1934–1946, 2012.

[29] M. Pierobon, I. Akyildiz, "Diffusion-based noise analysis for molecular communication in nanonetworks," IEEE Transactions on Signal Processing, vol. 59, no. 6, June 2011, pp. 2532–2547.

[30] M. Pierobon, I.F. Akyildiz, "Noise analysis in ligand-binding reception for molecular communication in nanonetworks", IEEE Transactions on Signal Processing, vol. 59, no. 9, September 2011, pp. 4168-4182.

[31] B. Atakan, O.B. Akan, "Deterministic capacity of information flow in molecular nanonetworks", Nano Communication Networks, vol. 1, no. 1, pp. 31–42, March 2010.

[32] N. Farsad, A.W. Eckford, S. Hiyama, Y. Moritani, "On-chip molecular communication: analysis

and design", IEEE Transactions on NanoBioscience, vol. 11, no. 3, 2012, pp. 304-314.

[33] K.V. Srinivas, A.W. Eckford, R.S. Adve, "Molecular communication in fluid media: the additive inverse Gaussian noise channel", IEEE Transactions on Information Theory, vol. 58, no. 7, July 2012, pp. 4678 - 4692.

[34] J.M. Jornet, I.F. Akyildiz, "Channel Modeling and Capacity Analysis for Electromagnetic Wireless Nanonetworks in the Terahertz Band," IEEE Transactions on Wireless Communications, vol. 10, no. 10, pp. 3211-3221, October 2011.

[35] I.F. Akyildiz, J.M. Jornet, "Electromagnetic Wireless Nanosensor Networks," Nano Communication Networks, Vol. 1, No. 1, pp. 3-19, March 2010.

[36] A. Guney, B. Atakan, O. B. Akan, "Mobile Ad Hoc Nanonetworks with Collision-based Molecular Communication," IEEE Transactions on Mobile Computing, vol. 11, no. 3, pp. 353-266, March 2012.

[37] J.N. Daigle, M. Femminella, Z. Shariat-Madar, "Modeling the spontaneous reaction of mammalian cells to external stimuli", in *Adhocnets 2012*, Springer, LNICST 111, J. Zheng et al. (Eds.), pp. 226–241, 2013.

[38] P. Sojan Lal, A. Unnikrishnan, K. Poulose Jacob, "Parallel implementation of octtree generation algorithm", International Conference on Image Processing (ICIP), 1998.

[39] R. Prodan, T. Fahringer, *Grid computing: experiment management, tool integration, and scientific workflows*, Springer-Verlag Berlin, LNCS 4340, 2007, ISBN 978-3-540-69261-4.

[40] GridGain, http://www.gridgain.com/

[41] Hans Hacker et al., "Considering GPGPU for HPC centers: is it worth the effort?", in *Facing the Multicore-Challenge: Aspects of New Paradigms and Technologies in Parallel Computing*, Springer-Verlag Berlin, LNCS 6310, 2010, R. Keller, D. Kramer, J.-P. Weiss (Eds.), pp. 118-121, ISBN 978-3-642-16233-6.

[42] S. Mlot, "Oak Ridge's Titan Named World's Fastest Supercomputer", Pcmag.com, November 12, 2012, http://www.pcmag.com/article2/0,2817,2412012,00.asp.

[43] M. Femminella, R. Francescangeli, G. Reali, H. Schulzrinne, "Gossip-based signaling dissemination extension for next steps in signaling", IEEE Network Operations and Management Symposium (NOMS), 2012, Maui, HW.

[44] M. Femminella, R. Francescangeli, F. Giacinti, E. Maccherani, A. Parisi, G. Reali, "Design, Implementation, and Performance Evaluation of an Advanced SIP-Based Call Control for VoIP Services, IEEE ICC '09, Dresden, Germany, 2009.

[45] B. Van Den Bossche, S. Van Hoecke, C. Danneels, J. Decruyenaere, B. Dhoedt, F. De Turck, "Design of a JAIN SLEE/ESB-based platform for routing medical data in the ICU", Computer Methods and Programs in Biomedicine, 2008, vol. 91, no. 3, pp. 265-277.

[46] M. Femminella, E. Maccherani, G. Reali, G., "Workflow Engine Integration in JSLEE AS", IEEE Communications Letters, vol. 15, no. 12, 2011, pp. 1405-1407.

[47] R. A. Freitas Jr., "Exploratory Design in Medical Nanotechnology: A Mechanical Artificial Red Cell," Artificial Cells, Blood Substitutes, and Immobil. Biotech., vol. 26, pp. 411–430, 1998.

[48] L. P. Giné, I. F. Akyildiz, " Molecular communication options for long range nanonetworks", Computer Networks, vol. 53, no. 16, 2009, pp. 2753–2766.

[49] Harvard University, "Toward Synthetic Life: Scientists Create Ribosomes -- Cell Protein Machinery", 9 March 2009, ScienceDaily, retrieved 9 May 2013, from http://www.sciencedaily.com- /releases/2009/03/090309104434.htm.

[50] B. Lewandowski et al., "Sequence-Specific Peptide Synthesis by an Artificial Small-Molecule Machine", Science, vol. 339, no. 6116, 11 January 2013, DOI: 10.1126/science.1229753, pp. 189-193.

[51] J. M. Jornet, I. F. Akyildiz, "Joint Energy Harvesting and Communication Analysis for Perpetual Wireless NanoSensor Networks in the Terahertz Band," IEEE Transactions on Nanotechnology, vol. 11, no. 3, pp. 570-580, May 2012.

[52] COMSOL Multiphysics®, available at http://www.comsol.com/

[53] H. ShahMohammadian, G.G. Messier, S. Magierowski, "Optimum receiver for molecule shift keying modulation in diffusion-based molecular communication channels," Nano Communication Networks, vol. 3, pp. 183–195, 2012.

[54] F. Beer, E.R. Johnston Jr., E. Eisenberg, P. Cornwell, "Vector Mechanics for Engineers: Dynamics", 9th Ed., 2009, McGraw-Hill, ISBN 978-0077295493.

[55] T.E. Mallouk, A. Sen, " Powering Nanorobots ", Scientific American, 2009, vol. 300, pp. 72 - 77, doi: 10.1038/scientificamerican0509-72.

[56] R. Parker, "California Scientists Attempt Nanomachines Against Arterial Plaque", Nanotech for Biotech, 14 June 2005, available at http://www.futurepundit.com/archives/002832.html.

[57] S.K. Sahoo, S. Parveen, J.J. Panda, "The present and the future of nanotechnology in human healt care", Nanomedicine: Nanotechnology, Biology, and Medicine, vol. 3, 2007, pp. 20-31.

[58] JSR 22: JAIN[TM] SLEE API Specification, available at http://www.jcp.org/en/jsr/detail?id=22.

## Appendix: Inelastic Collisions

A rigid body may be subject to elastic deformation but then it returns to its previous shape. The reciprocal directions, after a collision between two rigid bodies, are evaluated through the components of the velocity vector of both objects. In particular, the condition on the kinetic energy is useful to classify what kind of collision is observed. If only the momentum is conserved after the collision, but the kinetic energy may change, the collision is said "inelastic". This energy may be converted on potential energy by deformation or by heat. So the final energy (after collision) is a fraction of the initial energy, scaled by the restitution coefficient *e*. This coefficient is equal to 1 for the elastic collision case and is equal to 0 for the completely inelastic one, in which the two bodies are fused into a single one. Any value between 0 and 1 modulates the relative variation of kinetic energy before and after collision and the two bodies are kept apart as two separated entities.

The analysis can be made by using both an inertial reference system or a center of mass of a group of bodies (center of gravity). We describe the collision events through the latter approach. Let $\vec{v}_{cm}$ the velocity of the center of mass, we can express the velocity $\vec{v}_1$ and $\vec{v}_2$ of the two particles as a function of $\vec{v}_{cm}$, where $\vec{v}_1'$ and $\vec{v}_2'$ are their velocity relative to the center of mass and $\vec{v}_{cm}$:

$$\vec{v}_1 = \vec{v}_1' + \vec{v}_{cm} \tag{26}$$

$$\vec{v}_2 = \vec{v}_2' + \vec{v}_{cm} \tag{27}$$

The center of mass ensures that the momentum $\vec{p}'$ of the system is null:

$$\vec{p}' = m_1\vec{v}_1' + m_2\vec{v}_2' = \vec{0} \tag{28}$$

Let $\vec{v}_{1i}$ and $\vec{v}_{2i}$ the velocities of the particles before the collision and $\vec{v}_{1f}$ and $\vec{v}_{2f}$ those after the collision. We can introduce the momentum before and after collision:

$$\vec{p}_i' = m_1\vec{v}_{1i}' + m_2\vec{v}_{2i}' = \vec{0} \tag{29}$$

$$\vec{p}_f' = m_1\vec{v}_{1f}' + m_2\vec{v}_{2f}' = \vec{0} \tag{30}$$

This means that the center of mass sees the two particles moving towards itself (before the collision) and then it sees them moving away with their momentum equals and opposite after the collision:

$$\vec{p}_{1i}' = -\vec{p}_{2i}' \tag{31}$$

$$\vec{p}_{1f}' = -\vec{p}_{2f}' \tag{32}$$

For the general case, $\vec{p}_{1i}' \neq -\vec{p}_{1f}'$ and $\vec{p}_{2i}' \neq -\vec{p}_{2f}'$,

$$\vec{v}_{1f}' = \frac{(m_1 - e \cdot m_2)\vec{v}_{1i}' + m_2(1+e)\vec{v}_{2i}'}{m_1 + m_2} \tag{33}$$

$$\vec{v}_{2f}' = \frac{(m_2 - e \cdot m_1)\vec{v}_{2i}' + m_1(1+e)\vec{v}_{1i}'}{m_1 + m_2} \tag{34}$$

The kinetic energy after collision in the system of the center of mass results:

$$E_{kf}' = e^2 E_{ki}', \tag{35}$$

where $E_{ki}'$ and $E_{kf}'$ are given by

$$E_{kf}' = (\frac{1}{2}m_1 v_{1f}'^2 + \frac{1}{2}m_2 v_{2f}'^2) \tag{36}$$

$$E_{ki}' = (\frac{1}{2}m_1 v_{1fi}'^2 + \frac{1}{2}m_2 v_{2i}'^2) \tag{37}$$

For the case of an inelastic collision with the vessel surface, we can assume, in the previous equation, that the second particle has infinite mass and zero velocity, so we obtain:

$$\vec{v}'_{1f} = \frac{(m_1 - e \cdot m_2)\vec{v}'_{1i} + m_2(1+e)\vec{v}'_{2i}}{m_1 + m_2} \underset{v_{2i}=0}{=} \frac{(m_1 - e \cdot m_2)\vec{v}'_{1i}}{m_1 + m_2} = \lim_{m_2 \to \infty} \frac{\frac{m_1\vec{v}'_{1i}}{m_2} - e\vec{v}'_{1i}}{\frac{m_1}{m_2} + 1} = -e\vec{v}'_{1i} \qquad (38)$$

This means that the final velocity in the system of the center of mass is a fraction of the initial velocity and propagates in the opposite direction. As a matter of fact, the direction depends on the incident angle of the particle on the plane which is tangent to considered surface in the point of collision with the particle.